\documentclass[aps, pra, 10pt, twocolumn, superscriptaddress]{revtex4-1}
\pdfoutput=1

\usepackage{amsmath}
\usepackage{amssymb}
\usepackage{amsfonts}

\usepackage{graphicx}
\usepackage{float}

\usepackage{booktabs}
\usepackage{tabularx}
\usepackage{siunitx}
\usepackage{multirow}
\usepackage{tikz}
\usepackage{quantikz}

\usetikzlibrary{
    positioning,
    arrows.meta,
    shapes.geometric,
    decorations.pathreplacing,
    decorations.pathmorphing,
    calc,
    fit,
    backgrounds
}

\usepackage{listings}
\usepackage[dvipsnames]{xcolor}

\usepackage[colorlinks=true, citecolor=blue]{hyperref}

\makeatletter
\renewcommand{\p@subsection}{\thesection-}
\renewcommand{\p@subsubsection}{\thesection-\thesubsection}
\makeatother

\begin{document}

\title{Reduced Cost Quantum Kernel Training}
\author{Daniel R. Germain\textsuperscript{*}}
\email{drgermain@ncat.edu}
\affiliation{North Carolina A\&T, Greensboro, United States}

\author{Benjamin Goldweber\textsuperscript{*}}
\email{bensgoldweber@gmail.com}
\affiliation{Montgomery College, Bethesda, United States}

\author{Kayla J. Rodriguez\textsuperscript{*}}
\email{kjrodriguez1@ncat.edu}
\affiliation{North Carolina A\&T, Greensboro, United States}

\author{Kristen Rhinehardt}
\email{krinehardt@ncat.edu}
\affiliation{North Carolina A\&T, Greensboro, United States}

\begin{abstract}
Quantum kernel methods are promising for near-term quantum machine learning, but their training costs are driven by the quadratic scaling of Gram-matrix construction and the measurement cost of estimating individual kernel values. In this work, we investigate one-class anomaly detection as a resource-efficient operating regime for training non-variational quantum kernel models. Using the ULB Credit Card Fraud Detection dataset, we evaluate three quantum feature maps across training-set sizes and qubit counts. Power-law saturation models reveal strong diminishing returns with increasing training data, with 500-sample models retaining approximately 90--93\% of their fitted asymptotic average precision (AP). We further observe no severe exponential concentration among the evaluated feature maps
over the investigated qubit range,  thereby avoiding the exponentially increasing measurement requirements associated with that regime. Finally, Nystr\"om approximation reduces kernel construction from $\mathcal{O}(N^2)$ to
$\mathcal{O}(Nm)$ evaluations, with QSVR retaining 94--97\% of its
exact-kernel AP using only five landmarks. For ZZ-QSVR, downsampling from 4,000 to 500 training samples and applying a five-landmark approximation decreases the number of required kernel evaluations from approximately $8.0$ million to $2,485$, a reduction of more than 3,200-fold, while AP decreases from $0.812$ to $0.737$.
Under the minimum IonQ pricing assumption considered here, this corresponds to an illustrative reduction in quantum kernel training cost from approximately \$1.34 billion to \$417{,}480. These results demonstrate that training-set reduction and low-rank approximation can substantially reduce the resources required to train non-variational quantum kernel anomaly detectors while retaining meaningful anomaly-discrimination performance.
\end{abstract}

\maketitle
\section{Introduction}
With the rapid development of quantum computing, a wide range of methods and algorithms have been proposed to achieve quantum advantage~\cite{biamonte2017,liu2021,schuld2022}. In machine learning specifically, quantum kernel methods have emerged as a particularly promising approach. It has been shown that many quantum machine learning algorithms can be expressed in terms of quantum kernels~\cite{schuld2019,havlicek2019}, and this framework benefits from robust classical kernel theory and decades of industrial applications in classification, regression, and anomaly detection. Kernels have long been a workhorse of  machine learning, and in recent years, ample studies~\cite{mengoni2019kernel, tanner2026nonvariational} have suggested that kernels could benefit from quantum hardware.

Quantum approaches to support vector machines have been explored using quantum linear algebra methods~\cite{rebentrost2014,dunjko2018}, quantum feature maps~\cite{havlicek2019,schuld2019}, and variational quantum circuits~\cite{park2023vqasvm}. Quantum linear algebra approaches, such as the method proposed  in Ref.~\cite{rebentrost2014}, suggest exponential speedups for kernel matrix construction by exploiting quantum representations of data. Although promising in theory, these approaches generally rely on fault-tolerant quantum computers and substantial quantum resources, making them impractical for current noisy intermediate-scale quantum (NISQ) hardware.

\begin{figure}[t]
\centering
\includegraphics[
width=1.0\linewidth
]{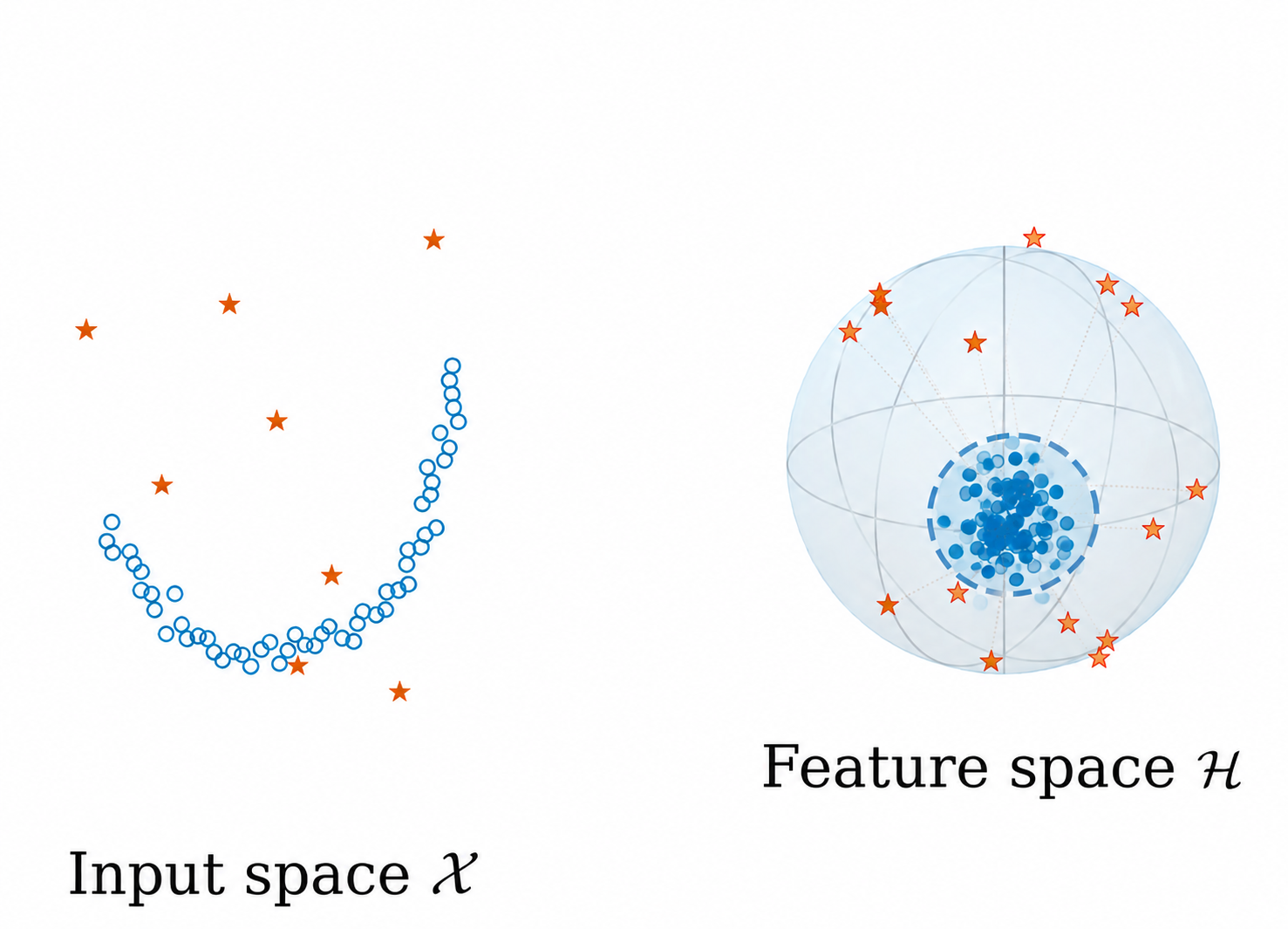}
\caption{Illustration of support vector data description (SVDD). Input data are mapped into a higher-dimensional feature space, where the nominal samples are enclosed by a minimum-volume hypersphere and points outside the boundary are identified as anomalies.}
\label{fig:svdd}
\end{figure}

An alternative approach is to use quantum computation not to accelerate the classical construction of the kernel, but to directly define and evaluate the kernel through a quantum feature map. Quantum feature map approaches encode classical data into quantum states and use measurements of those states to estimate similarities between data points~\cite{schuld2019,havlicek2019}. In this setting, the quantum computer serves as part of the kernel evaluation procedure rather than as a subroutine for accelerating a conventional linear-algebra algorithm. Because these methods can be implemented using relatively shallow circuits and modest qubit counts, they are particularly attractive for near-term quantum hardware~\cite{preskill2018}. However, several fundamental challenges limit their practical scalability.

\begin{enumerate}
    \item First, constructing a quantum kernel matrix requires $\mathcal{O}(N^2)$ pairwise evaluations~\cite{schuld2019, havlicek2019}. For 500 training samples, the Gram matrix contains 250,000 entries but only 124,750 unique off-diagonal for a symmetric kernel. Evaluating these values on quantum hardware requires circuit executions that can make kernel construction prohibitively costly.
    
    \item Second, quantum kernels can suffer from exponential concentration, 
    where kernel values become increasingly similar and the kernel matrix approaches an uninformative form (such as a quasi-identity matrix)~\cite{thanasilp2024}. Resolving meaningful differences between these concentrated values requires exponentially more measurement shots; for example, a strongly concentrated 20-qubit kernel may require on the order of $10^6$ measurement shots per kernel calculation~\cite{thanasilp2024}. 
    \item Third, effective quantum feature-map design remains challenging. No general prescription ensures that a feature map avoids exponential concentration~\cite{thanasilp2024}, captures useful data structure~\cite{schuld2019}, and consistently outperforms classical kernels. Performance varies across datasets, making feature-map selection a problem-specific design task. 
\end{enumerate}

Variational quantum kernels have been proposed as a means of addressing the feature map design problem by using a parameterized quantum circuit (PQC) to learn an effective quantum feature map. However, this approach does not eliminate the computational cost associated with kernel evaluation and can introduce barren plateaus, a phenomenon in which gradients vanish exponentially as the size of the system increases~\cite{mcclean2018barren}.

These limitations motivate the consideration of learning problems that can operate effectively with small training sets, since reducing the number of training samples directly reduces the number of required quantum kernel evaluations. An additional consideration is the geometry of the training data. If the samples remain sufficiently close after being embedded in quantum Hilbert space, their pairwise fidelities can remain large enough to mitigate global-measurement-induced exponential concentration~\cite{thanasilp2024}. As discussed above, mitigating exponential concentration is particularly important for NISQ implementations, as it can reduce the measurement-shot overhead required to approximate quantum kernel values on hardware.

One-class anomaly detection provides a promising setting in which to investigate these resource requirements. One-class methods learn only the nominal behavior of a system and therefore require only nominal samples during training. This removes the requirement for labeled anomalous examples and provides a training population drawn from a single class, which may offer favorable data geometry for fidelity-based quantum kernels. Combined with the potential for one-class models to operate using relatively small nominal training sets, these properties motivate their use in this work as a candidate regime for investigating the economic feasibility of training quantum kernels.

In this work, we show that quantum kernel training is economically feasible by addressing three primary sources of quantum resource cost.  First, we evaluate the sensitivity of anomaly-detection performance to the number of nominal training samples used to construct the quantum Gram matrix (Sec.~\ref{sec:downsample}), determining how aggressively the training set can be reduced without substantial loss in average precision. Second, we analyze exponential concentration across increasing qubit counts and feature-map architectures to determine whether the resulting fidelity values remain measurable with practical shot budgets (Sec.~\ref{sec:kernel-concentration-analysis}). Finally, we apply the Nystr\"om approximation~\cite{williams2001nystrom} to reduce kernel construction from $\mathcal{O}(N^2)$ to $\mathcal{O}(Nm)$ evaluations, where $m \ll N$ is the number of landmark samples, and quantify the corresponding trade-off in anomaly-detection
performance (Sec.~\ref{sec:nystrom-approx-results}).

The results demonstrate that these strategies can reduce quantum
kernel-training requirements by several orders of magnitude while retaining meaningful anomaly-discrimination performance. For the ZZ-QSVR configuration, the largest experimentally evaluated training set contains 4{,}000 nominal samples and achieves an AP of $0.812$. Exploiting kernel symmetry and the known unit diagonal, the corresponding exact Gram matrix requires 7{,}998{,}000 unique quantum kernel evaluations. Downsampling to 500 nominal samples and applying a five-landmark Nystr\"om approximation reduces this requirement to approximately 2{,}485 evaluations, corresponding to a greater
than 3{,}200-fold reduction.

At IonQ's minimum estimated cost of \$168 per kernel
evaluation~\cite{ionq2025}, this experimentally grounded reduction
corresponds to an illustrative kernel-training cost decrease from
approximately \$1.34 billion to \$417{,}480. For ZZ-QSVR, the associated AP
decreases from $0.812$ to $0.737$, corresponding to a relative reduction of
approximately $9.2\%$. Moreover, the fitted power-law model predicts an
asymptotic AP of only $0.8401$ for ZZ-QSVR, such that the aggressively
reduced model remains within approximately $12.3\%$ of the predicted
infinite-sample limit. Thus, even relative to the performance expected from
substantially larger training sets, the proposed resource reductions retain
most of the attainable AP. The evaluated quantum kernels also remain outside
the severe exponential-concentration regime over the investigated range of
1--20 qubits, avoiding the infeasible shot requirements associated with resolving severely concentrated fidelity values.

\begin{table*}[t]
\centering
\caption{Examples of classical kernels, their feature-space dimensionality, and their similarity interpretations. Here $n$ denotes the input dimensionality and $d$ the polynomial degree.}
\label{tab:classical_kernels}
\vspace{2mm}
\small
\renewcommand{\arraystretch}{1.4}
\setlength{\tabcolsep}{8pt}
\begin{tabular}{l c c c}
\toprule[0.9pt]

\textbf{Kernel} & \textbf{Kernel Function} & \textbf{Hilbert Space Dim.} & \textbf{Similarity Interpretation} \\
\midrule[0.4pt]

Linear & $k(x,x') = x^{\top}x'$ & $n$ & Vector alignment \\
Polynomial & $k(x,x') = (x^{\top}x' + c)^d$ & $\binom{n+d}{d}$ & Higher-order feature interactions \\
Gaussian (RBF) & $k(x,x') = \exp\!\left(-\gamma \lVert x-x'\rVert^2\right)$ & $\infty$ & Distance-based similarity \\
\bottomrule[0.9pt]
\end{tabular}
\end{table*}

\section{Background}
\subsection{Quantum and Classical Kernels}\label{sec:quantum-kernels}
Kernels can be thought of as functions that capture a notion of similarity between data points, often in a higher-dimensional feature space. Mathematically, given an input domain $\mathcal{X}$, and a feature map $\Phi$ that projects inputs $x$ into a  \textit{Hilbert space} $\mathcal{H}$ (or the feature space), 

\begin{equation}
\Phi: \mathcal{X} \rightarrow \mathcal{H},
\qquad
x \mapsto \Phi(x)
\label{eq:feature_map}
\end{equation}

\noindent the kernel function, $k(x,x') $, is defined as the inner product between the two mapped feature vectors, $\Phi(x)$ and $\Phi(x')$.

\begin{equation}
k(x,x')=\langle \Phi(x),\Phi(x')\rangle_{\mathcal H}
\label{eq:kernel}
\end{equation}

\noindent If a closed-form kernel function equivalent to a higher-dimensional inner product can be found, then a higher-dimensional notion of similarity can be calculated without explicitly constructing the feature vectors $\Phi(x)$ and $\Phi(x')$. 
This observation, known as the \emph{kernel trick}~\cite{aizerman1964,boser1992}, allows machine learning algorithms to implicitly operate in high-dimensional (or even infinite-dimensional) feature spaces while avoiding the computational cost of explicitly computing higher-dimensional feature representations.
A few common kernels and their corresponding closed-form kernel functions are summarized in Table~\ref{tab:classical_kernels}. Each kernel induces a different feature space and, therefore, encodes a different notion of similarity between data points. Because the kernel defines the inner products in the feature space, it also determines the geometry in which the data is represented, influencing the decision boundaries and patterns that kernel-based learning algorithms can identify~\cite{scholkopf2002}.

\textit{Quantum kernels} extend the classical kernel framework by replacing the classical feature map with a quantum feature map~\cite{havlicek2019,schuld2019}. A quantum state $|\psi\rangle$ is represented by a vector in a complex-valued Hilbert space. Consequently, encoding classical data into a quantum state naturally defines a feature map: each classical input $x$ is mapped to a  quantum state vector $|\psi(x)\rangle$ in the Hilbert space of the quantum system. For a system of $n$ qubits, the Hilbert space has dimension $2^n$, and the quantum feature map can be expressed as the mapping of the input domain to the $2^n$-dimensional Hilbert space

\begin{equation}
\Phi_q: \mathcal{X} \rightarrow \mathcal{H}_{2^n},
\qquad
x \mapsto |\psi(x)\rangle,
\label{eq:quantum-feature-map}
\end{equation}

Once the inputs are mapped into this quantum feature space, a kernel can be constructed by measuring the similarity between quantum states. In this work, \textbf{fidelity-based} quantum kernels are considered~\cite{havlicek2019}, for which the kernel is defined as the squared magnitude of the inner product between two encoded states:

\begin{equation}\label{eq:fidelity-def}
k(x,x') =
\left|
\langle \psi(x) \mid \psi(x') \rangle
\right|^2 ,
\end{equation}

\noindent where $k(x,x')$ corresponds to the fidelity between the two pure quantum states and takes values between zero and one. Values closer to one indicate greater similarity between the corresponding quantum states, while values closer to zero indicate that the states have little overlap. The resulting kernel therefore defines a similarity measure based on \textit{quantum state overlap}.

\subsection{Quantum Feature Maps}\label{sec:quantum-feature-maps}
The quantum feature map determines how the classical input data is encoded in the quantum Hilbert space and therefore defines the properties of the resulting quantum kernel~\cite{havlicek2019,schuld2019}. In general, a quantum feature map can be represented by a parameterized quantum circuit (PQC)~\cite{benedetti2019pqc,cerezo2021vqa} that acts on an initial reference state. Although other initial states may be used, quantum circuits are commonly initialized in the all-zero computational basis state, $|0\rangle^{\otimes n}$, where $n$ is the number of qubits.

The quantum feature map can therefore be represented as a unitary transformation $U_{\Phi}(x,\boldsymbol{\theta})$ that acts on this initial reference state to produce the data-encoded quantum state vector: 

\begin{equation}
|\psi(x)\rangle =
U_{\Phi}(x,\boldsymbol{\theta})
|0\rangle^{\otimes n},
\label{eq:general_qfm}
\end{equation}

\noindent where $x$ denotes the classical input, $\boldsymbol{\theta}$ represents any additional circuit parameters, and $U_{\Phi}$ is the unitary operator implementing the quantum feature map. 

\begin{figure}[h]
\centering
\includegraphics[width=0.8\linewidth]{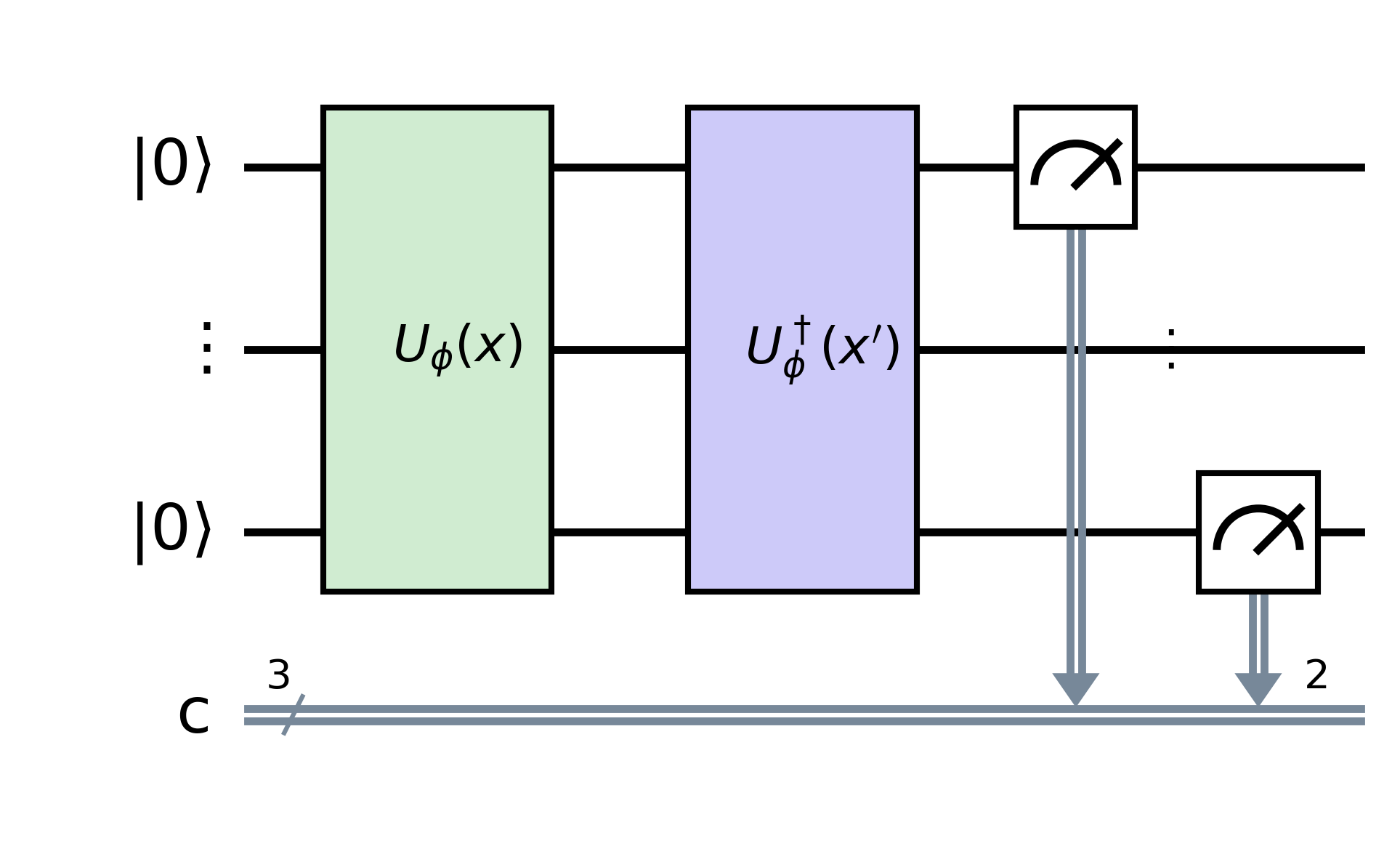}
\[\Downarrow \text{ repeat for all } (x_i, x_j) \in \mathcal{X}\]
\vspace{0.3cm}
$K =
\begin{pmatrix}
\kappa(x_1,x_1) & \kappa(x_1,x_2) & \cdots & \kappa(x_1,x_m) \\
\kappa(x_2,x_1) & \kappa(x_2,x_2) & \cdots & \kappa(x_2,x_m) \\
\vdots          & \vdots          & \ddots & \vdots          \\
\kappa(x_m,x_1) & \kappa(x_m,x_2) & \cdots & \kappa(x_m,x_m)
\end{pmatrix}$
\caption{Quantum kernel evaluation circuit and Gram matrix
construction. For each pair of data points $(x_i, x_j)$, the
quantum kernel $\kappa(x_i, x_j) = |\langle 0|^{\otimes N}
\mathcal{U}_\phi^\dagger(x_j)\mathcal{U}_\phi(x_i)|0\rangle^{\otimes N}|^2$
is estimated as the return probability of the all-zero bitstring.
Repeating this procedure for all pairs of input samples assembles the Gram matrix
$K$, which is passed directly to the classical anomaly
detection solver.}
\label{fig:kernel_gram}
\end{figure}

It is worth noting that not all quantum feature maps contain trainable parameters, and the feature maps considered in this work are non-variational. Aside from the input-dependent parameters used to encode $x$, any additional circuit parameters are fixed prior to evaluation. Using the fidelity kernel defined in Eq.~\ref{eq:fidelity-def}, the overlap between two states encoded by the same non-variational feature map can then be written as

\begin{equation}
K(x,x') =
\left|
\langle 0|^{\otimes n}
U_{\Phi}^{\dagger}(x')
U_{\Phi}(x)
|0\rangle^{\otimes n}
\right|^2.
\label{eq:quantum_kernel}
\end{equation}

\noindent and the resulting kernel fidelity can be interpreted as the probability of measuring the all-zero computational basis state, $\ket{0}^{\otimes n}$. Repeating this evaluation for every pair of samples assembles the Gram matrix, as illustrated in Fig.~\ref{fig:kernel_gram}. The quantum feature maps considered in this work are further described in Section \ref{sec:feature-map-selection}.

\begin{figure*}[t]
    \centering
    \includegraphics[width=0.7\linewidth]{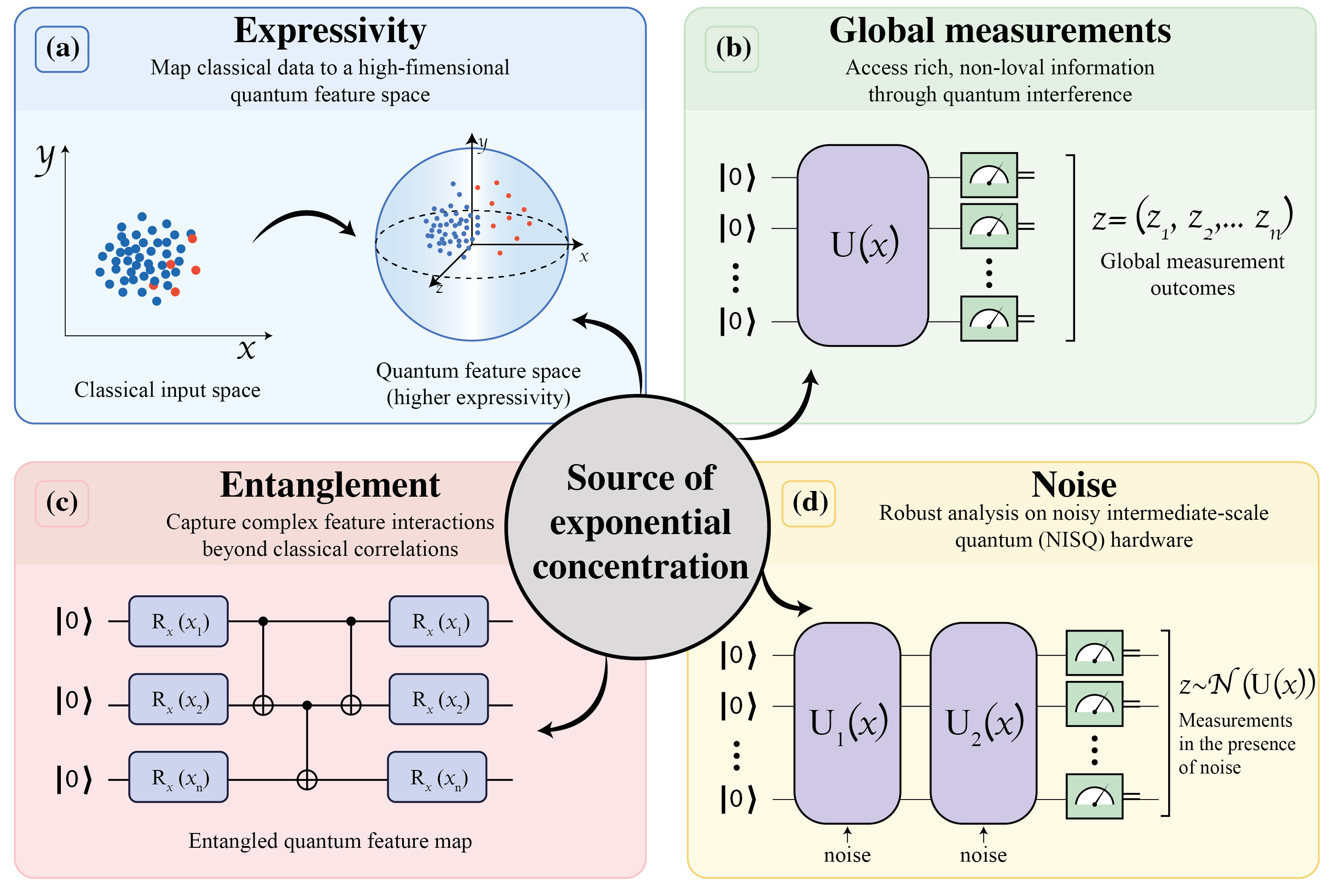}
    \caption{Illustration of the four primary mechanisms associated with exponential concentration in quantum kernels: (a) expressivity-induced concentration, arising as increasingly expressive feature maps approach highly expressive state ensembles; (b) global-measurement-induced concentration, associated with global overlap measurements over an exponentially growing Hilbert space; (c) entanglement-induced concentration, resulting from increasing entanglement between measured and unmeasured subsystems; and (d) noise-induced concentration, arising from the accumulation of hardware noise with increasing circuit depth and system size.}
    \label{fig:concentration-mechanisms}
\end{figure*}

\subsection{Semi-Supervised Kernel Anomaly Detectors}
With a feature map and corresponding kernel function defined, the
similarity between pairs of data points can be evaluated in the resulting
feature space. Given a dataset $\mathcal{X}=\{x_1,\ldots,x_m\}$, evaluating
the kernel between every pair of samples produces a pairwise similarity
matrix $K$, whose entries are given by
\begin{equation}
K_{ij} = k(x_i,x_j)
\label{eq:gram_matrix}
\end{equation}
where $k$ is the kernel as defined in Eq.\ref{eq:kernel}. For many of the learning guarantees developed in classical kernel theory to hold, the resulting matrix must be symmetric such that

\begin{equation}
K_{ij} = K_{ji},
\label{eq:kernel_symmetry}
\end{equation}
and positive semi-definite
\begin{equation}
\mathbf{z}^{T}\mathbb{K}\mathbf{z} \geq 0
\qquad
\forall\,\mathbf{z}\in\mathbb{R}^{m}.
\label{eq:kernel_psd}
\end{equation}

\noindent Matrices satisfying these conditions are called \textit{Gram matrices}~\cite{mercer1909,scholkopf2002}, and they are discussed further in Appendix~\ref{sec:one_class_anomalies}. Noiseless quantum fidelity kernels always produce valid gram matrices.

Many kernel methods formulate their optimization problems entirely in terms of the Gram matrix~\cite{scholkopf2002}, allowing them to operate on the geometry induced by the kernel without explicitly constructing the corresponding feature-space representation. The kernel-based methods considered in this work are the one-class support vector machine (OC-SVM) and support vector regression (SVR). While OC-SVM is explicitly designed for one-class classification, SVR is conventionally a regression method. Here, SVR is adapted for anomaly detection using a reconstruction-error approach, in which deviations from the learned nominal behavior are used to identify anomalous samples. Both methods are described in greater detail in the Appendix~\ref{sec:one_class_anomalies}.

To quantum-enhance these methods, the classical Gram matrix is replaced by one generated using a quantum kernel. The optimization procedure then operates directly on the geometry induced by the quantum feature map, without requiring explicit access to the underlying quantum states. This produces the quantum-enhanced methods considered in this work, denoted QOC-SVM and QSVR~\cite{osvm_fraud,tscharke2025nisq}.

We evaluate both anomaly detection methods to determine whether their differing optimization objectives and uses of the kernel matrix result in different sensitivities to reductions in the number of training samples. 

\subsection{Nystr\"om Method}
\label{sec:nystrom-method}

The Nystr\"om method provides a low-rank approximation to a kernel Gram
matrix using a subset of the training samples, referred to as
\textit{landmark points}. Consider a training set
$\mathcal{X}=\{\mathbf{x}_1,\ldots,\mathbf{x}_N\}$ with Gram matrix
$K\in\mathbb{R}^{N\times N}$, where

\begin{equation}
K_{ij}=k(\mathbf{x}_i,\mathbf{x}_j).
\end{equation}

Rather than evaluating kernel fidelity between each pair of training samples,
Nystr\"om selects a subset of $m\ll N$ landmark points,

\begin{equation}
\mathcal{L}
=
\{\mathbf{x}_{\ell_1},\ldots,\mathbf{x}_{\ell_m}\}
\subset\mathcal{X}.
\end{equation}

\noindent The kernel is then evaluated between each training sample and each landmark,
forming the cross-kernel matrix

\begin{equation}
C = K_{N,m}\in\mathbb{R}^{N\times m},
\qquad
C_{ij}=k(\mathbf{x}_i,\mathbf{x}_{\ell_j}).
\label{eq:nystrom-cross}
\end{equation}

The landmark-to-landmark kernel matrix is

\begin{equation}
W = K_{m,m}\in\mathbb{R}^{m\times m},
\qquad
W_{ij}=k(\mathbf{x}_{\ell_i},\mathbf{x}_{\ell_j}).
\label{eq:nystrom-landmark}
\end{equation}

Because the landmarks are selected from the training set, $W$ is a
submatrix of $C$ and therefore does not require an independent set of
kernel evaluations. The full Gram matrix is then approximated as

\begin{equation}
\widetilde{K}
=
CW^{\dagger}C^{T},
\label{eq:nystrom-approx}
\end{equation}

\noindent where $W^{\dagger}$ denotes the Moore--Penrose pseudoinverse of
the landmark kernel matrix. If $W$ is nonsingular, $W^{\dagger}=W^{-1}$.
The resulting approximation
$\widetilde{K}\in\mathbb{R}^{N\times N}$ therefore has the same dimensions
as the original Gram matrix, but is reconstructed from kernel evaluations
involving only the $m$ selected landmarks. Its rank is at most $m$.

For the quantum kernels considered in this work, each entry of $C$ is
obtained using the same fidelity-kernel circuit employed for exact Gram
matrix construction,

\begin{equation}
C_{ij}
=
\left|
\langle 0|^{\otimes n}
U_{\Phi}^{\dagger}(\mathbf{x}_{\ell_j})
U_{\Phi}(\mathbf{x}_i)
|0\rangle^{\otimes n}
\right|^2.
\end{equation}

Thus, Nystr\"om approximation does not modify the underlying quantum kernel
or the circuit used to evaluate it. Instead, it reduces the number of
sample pairs for which the quantum circuit must be executed, with the
remaining entries of the approximate Gram matrix reconstructed
classically through Eq.~\ref{eq:nystrom-approx}.

For a symmetric kernel with known unit diagonal, an exact $N$-sample Gram
matrix requires $N(N-1)/2$ unique off-diagonal kernel evaluations. With
$m$ landmarks selected from the training set, the Nystr\"om construction
requires

\begin{equation}
m(N-m)+\frac{m(m-1)}{2}
\end{equation}

\noindent unique nontrivial kernel evaluations when symmetry and the known
diagonal are exploited. The dominant scaling is therefore
$\mathcal{O}(Nm)$ rather than $\mathcal{O}(N^2)$ for $m\ll N$. For example,
with $N=500$ and $m=5$, the exact Gram matrix requires 124{,}750 unique
off-diagonal kernel evaluations, whereas applying the Nystr\"om approximation
requires only 2{,}485. The resulting approximate Gram matrix nevertheless
remains $500\times500$ and can be supplied directly to the classical
kernel-based anomaly-detection method.

\subsection{Exponential Concentration}
\label{sec:concentration}
Exponential concentration is a fundamental challenge for quantum kernel methods in which the distribution of kernel values becomes increasingly concentrated as the number of qubits grows~\cite{thanasilp2024}. In this regime, differences between kernel values can become exponentially small in the number of qubits, making the similarity between distinct samples
increasingly difficult to resolve under finite-shot measurement.

For fidelity kernels, exponential concentration can cause the off-diagonal kernel values to rapidly approach zero as the number of qubits increases, while the diagonal values remain unity. Consequently, the Gram matrix becomes increasingly identity-like. In particular, Proposition~3 of
Thanasilp \textit{et al.}~\cite{thanasilp2024} shows that, for a non-entangling product-state embedding under the input-distribution assumptions considered in their analysis, the fidelity kernel concentrates around

\begin{equation}
\mu(n)=2^{-n}.
\label{eq:fidelity_concentration_scale}
\end{equation}

\noindent At $n=20$ qubits, this characteristic scale is

\begin{equation}
2^{-20}\approx9.54\times10^{-7},
\end{equation}

\noindent illustrating that exponentially concentrated kernel values can reach the order of $10^{-6}$ even at the system sizes considered in this work. Resolving differences between fidelities at this scale with finite-shot measurements can require prohibitively large measurement budgets. Since each additional shot corresponds to an additional circuit execution on quantum hardware, exponential concentration can dramatically increase the computational and economic cost of kernel estimation. Addressing this behavior is therefore an important consideration for the practical implementation of quantum kernel methods.

To characterize this behavior, concentration is evaluated using the distribution of off-diagonal kernel values,

\begin{equation}
\mathcal{K}_{\mathrm{off}}
=
\left\{
\kappa(x_i,x_j)
\,\middle|\,
i\neq j
\right\},
\end{equation}

\noindent with particular emphasis on the median off-diagonal fidelity,

\begin{equation}
p_{50}(n)
=
\operatorname{median}_{i\neq j}
\left[
\kappa^{(n)}(x_i,x_j)
\right],
\label{eq:p50_concentration}
\end{equation}

\noindent where $n$ denotes the number of encoded features and corresponding qubits. The mean, $p_{95}$, and off-diagonal variance are additionally used to characterize the location and spread of the kernel-value distribution. Examining these quantities as $n$ increases provides an empirical measure of whether pairwise fidelities progressively collapse toward an exponentially small regime.

Thanasilp \textit{et al.}~\cite{thanasilp2024} identify four principal mechanisms through which exponential concentration can arise: expressivity, entanglement, hardware noise, and global measurement, as illustrated in Fig.~\ref{fig:concentration-mechanisms}. These mechanisms are considered separately below to establish which sources are present or controlled in the experimental setting considered in this work. 

\subsubsection{Entanglement-Induced Concentration}\label{sec:entanglement-induced-concentration}
Entanglement can independently induce concentration for certain classes of quantum kernels~\cite{thanasilp2024}. However, the entanglement-induced concentration mechanism identified in Thanasilp \textit{et al.}~\cite{thanasilp2024} does not independently apply to the fidelity kernels considered in this work, so it is not treated as an independent source of concentration in this analysis.

\subsubsection{Expressivity-Induced Concentration}\label{sec:expressivity-induced-concentration}
Highly expressive quantum feature maps can produce concentration by mapping data into increasingly complex quantum feature-space geometries, causing pairwise kernel values to become increasingly difficult to distinguish~\cite{thanasilp2024}. The expressive capacity of a quantum feature map can be influenced by several aspects of its architecture, including the number of qubits, circuit depth, data re-uploading, and the use of entangling operations. Although entanglement-induced concentration is not treated as an independent concentration mechanism for the fidelity kernels considered here, entangling operations can still contribute to concentration indirectly by increasing the expressive capacity and altering the geometry of the feature map. To investigate the impact of expressivity on exponential concentration, the feature maps considered in this work span increasing levels of circuit complexity and expressive capacity.

\subsubsection{Global-Measurement-Induced Concentration}\label{sec:global-measurement-induced-concentration}
\label{sec:exponential_concentration}
Independent of feature-map expressivity, global fidelity measurements can themselves induce exponential concentration. Even when the encoded feature-space geometry is generated by a simple, low-expressivity feature map with no entanglement, measuring similarity across the complete quantum state can cause kernel values to become exponentially small as the number of qubits increases~\cite{thanasilp2024}. Intuitively, this occurs because the global fidelity measurement compresses the relationship between two states in an exponentially growing Hilbert space into a single overlap value. As the number of qubits increases, this global overlap can become increasingly small, making differences between encoded states progressively more difficult to resolve with a finite number of measurements.

While posing a challenge for fidelity kernels, global-measurement-induced concentration can be mitigated when encoded data points remain sufficiently close in quantum Hilbert space~\cite{thanasilp2024}. This property is particularly relevant to the one-class setting considered in this work, where the training set consists exclusively of nominal samples exhibiting similar behavior. A suitably selected quantum feature map may preserve this similarity during encoding, maintaining proximity between the resulting quantum states and thereby reducing susceptibility to global-measurement-induced concentration.

\subsubsection{Noise-Induced Concentration}\label{sec:noise-induced-concentration}
Hardware noise provides an additional mechanism through which quantum kernel values can concentrate. Decoherence, gate errors, and measurement errors progressively degrade the encoded quantum states and can cause kernel values to converge toward noise-dependent limiting values as circuit depth increases~\cite{thanasilp2024}. The analysis in this work is performed using exact noiseless state-vector simulation. Hardware noise is therefore excluded from this analysis. 

\section{Methodology}\label{sec:methodolgy}
This section describes the sampling procedure, quantum feature maps, and experimental framework used in this study. The analysis is performed on the ULB Credit Card Fraud Detection dataset~\cite{kaggle_fraud}, which was selected in part because quantum one-class support vector machine (QOC-SVM) methods have previously been studied on this dataset~\cite{osvm_fraud}.

To evaluate kernel behavior as the number of qubits increases, PCA is used to generate input representations ranging from 1 to 20 components. For all feature maps considered in this work, each input feature is encoded on a single qubit; therefore, the PCA sweep directly produces quantum kernels ranging from 1 to 20 qubits. At each point in the sweep, 12 independent training sets are generated using different random seeds, resulting in 12 separate kernel matrices for each PCA step. Reported results are aggregated across these 12 kernels.

All quantum kernels were simulated using NVIDIA RTX 5080 and RTX 5090 GPUs. The resulting kernel matrices were provided as precomputed kernels to the anomaly-detection methods considered in this study, OC-SVM and reconstruction-error SVR, which were implemented using scikit-learn.

\subsection{Downsampling Methodology}
Due to the computational cost of quantum kernel evaluation and the
$\mathcal{O}(N^2)$ scaling of full kernel matrix construction, reducing the number of training samples substantially decreases the resources required for training kernel construction. To investigate the sensitivity of quantum kernel anomaly detection to training-set size, we first construct a fixed test set of 2,000 samples and then systematically downsample the remaining nominal data. Training-set sizes of 500, 1,000,
2,000, and 4,000 samples (denoted by $n_{\mathrm{train}}$) are considered.

The sampling procedure consists of two stages. First, random sampling is
used to construct the 2,000-sample test set with a fixed anomaly fraction of $5\%$. This is consistent with the experimental setting used by Kyriienko \textit{et al.}~\cite{osvm_fraud}. These samples are then removed from the full dataset, ensuring that no test sample can appear in any subsequent training set. The test set is generated only once and remains \textbf{identical across all training-set sizes, random seeds, anomaly-detection methods, and quantum feature maps.} This ensures that every configuration is evaluated using the same observations and class distribution, allowing performance differences to be attributed to changes in the training configuration rather
than variation in the test data.

After removing the test samples, the remaining nominal samples are clustered using $K$-means~\cite{lloyd1982}. Training samples are then drawn proportionally from each cluster to preserve the distribution of the nominal data while constructing a training set of size  $n_{\mathrm{train}}$. This procedure is repeated independently across the 12 random seeds introduced above, producing 12 sampled training sets at each value of $n_{\mathrm{train}}$. Overlap between training sets generated from different seeds is permitted, as the purpose of the repeated sampling is to quantify variability introduced by the downsampling procedure while maintaining the same underlying nominal population.

\subsection{Feature Map Selection}\label{sec:feature-map-selection}
As discussed in Section~\ref{sec:concentration}, the exponential concentration induced by noise and entanglement is not present in the simulated fidelity kernels considered in this work, leaving global measurement and expressivity as the two remaining sources of exponential concentration. The feature maps are therefore selected  with progressively increasing expressivity, beginning with a low-expressivity non-entangling feature map to isolate the effects of global measurement concentration. In this work, the feature maps considered are simple angle-encoding (Sec. \ref{sec:angle-encoding}), Full ZZ (Sec. \ref{sec:full-zz}), and Full ZZ with three layers of data re-uploading (Sec. \ref{sec:full-zz-reupload}).

To accommodate the different rotation structures and sensitivities of each feature map, a common parameterized scaling framework is applied to the standardized input features. For each feature $z_i$, the encoded value is defined as

\begin{equation}\label{eq:encoded-value}
x_i = \frac{\pi}{2} + \eta z_i,
\end{equation}

\noindent where $\eta$ is a feature-map-specific scaling parameter that controls the amount of standardized feature variation retained in the encoded rotation angles. Using the same scaling form with independently selected values of $\eta$ provides a flexible preprocessing framework across the feature maps, while allowing the input range to be adapted to the rotation structure of each encoding.

\subsubsection{Angle Encoding Feature Map}\label{sec:angle-encoding}

Angle encoding serves as the low-expressivity control for concentration analysis. The circuit consists only of independent single-qubit rotations and contains no entangling operations, resulting in a comparatively simple quantum embedding scheme ~\cite{schuld2019,mitarai2018}. Since hardware noise and entanglement-induced concentration are excluded by the experimental setting, the limited expressivity of angle encoding minimizes the remaining contribution from expressivity-induced concentration. Consequently, any concentration observed with this feature map provides the clearest setting to examine the effects of global measurement.

\begin{figure}[H] \centering \includegraphics[width=1.0\linewidth]{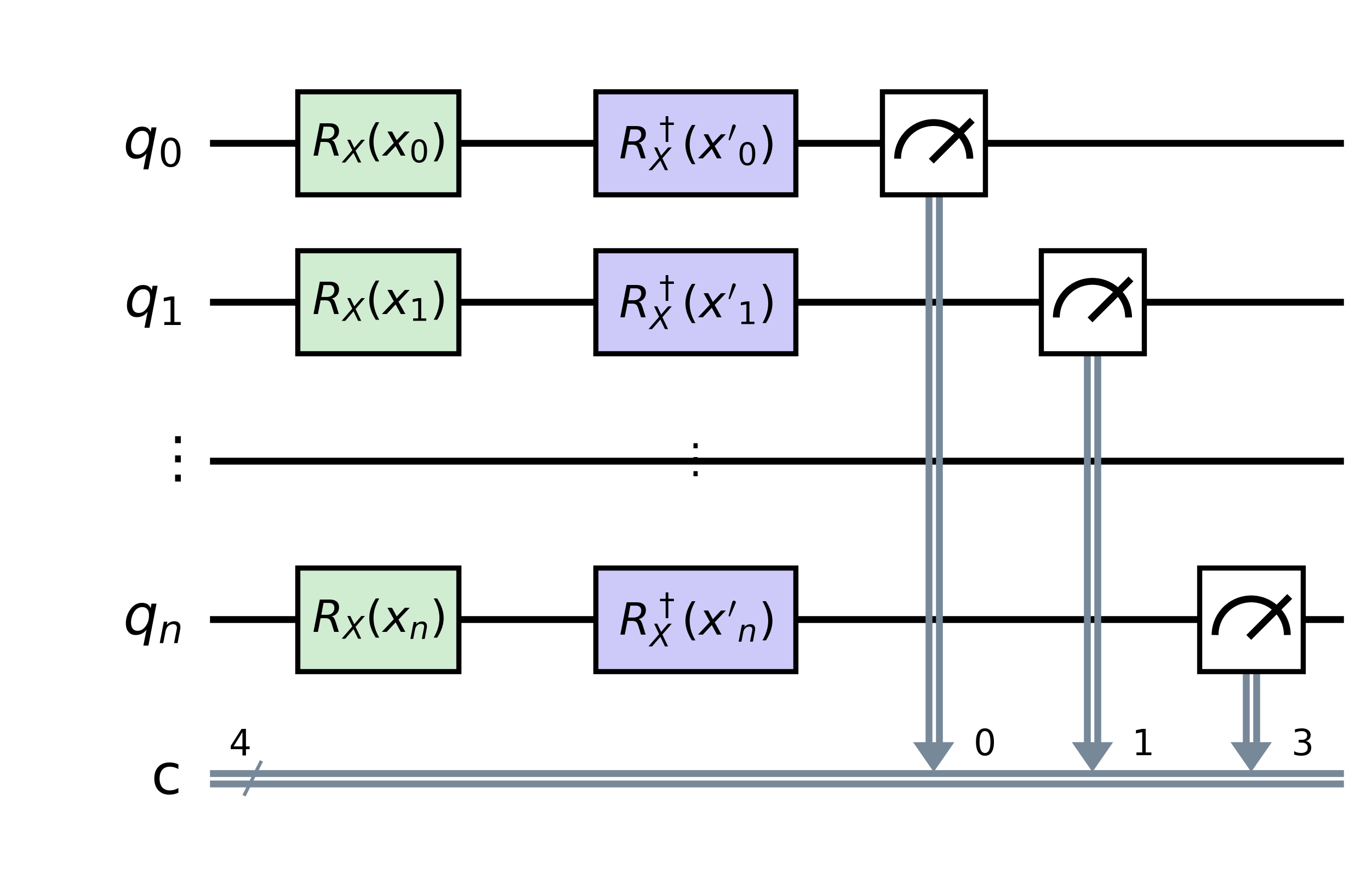} 
\caption{Angle-encoding quantum feature map. Each classical input feature $x_i$ is mapped directly to a corresponding qubit through a single-qubit rotation $R_X(x_i)$. The circuit begins with all qubits initialized in the $\lvert 0\rangle$ state, after which the feature-dependent rotations encode the classical input into the quantum state. No entangling gates are applied, such that each feature is encoded independently. The resulting quantum state is used to evaluate the fidelity kernel by measuring the overlap between states prepared from different input samples.}\label{fig:angle_encoding} \end{figure}

In angle encoding, each classical feature $x_i$ is encoded as a rotation angle on the corresponding qubit (Fig.~\ref{fig:angle_encoding}). The qubits are initialized in the computational basis state $|0\rangle^{\otimes n}$ and transformed using single-qubit rotations $R_X$,

\begin{equation}
U_{\mathrm{AE}}(x)
=
\bigotimes_{i=1}^{n} R_X(x_i),
\label{eq:angle_encoding}
\end{equation}

\noindent where

\begin{equation}
R_X(\theta)
=
\exp\left(-i\frac{\theta}{2}X\right),
\label{eq:rx_gate}
\end{equation}

\noindent and $X$ denotes the Pauli-$X$ operator. Because the circuit contains no multi-qubit operations, the resulting quantum state is a separable product of rotations applied to the computational ground state:

\begin{equation}
|\psi(x)\rangle
=
\bigotimes_{i=1}^{n} R_X(x_i)|0\rangle
\label{eq:angle_state}
\end{equation}

For the fidelity kernel defined in Eq.~\ref{eq:quantum_kernel}, the overlap between two angle-encoded states factorizes across qubits. Using

\begin{equation}
\langle 0|
R_X^\dagger(x_i)R_X(x_i')
|0\rangle
=
\cos\left(\frac{x_i-x_i'}{2}\right),
\label{eq:angle_overlap}
\end{equation}

\noindent the resulting kernel has the closed-form expression:

\begin{equation}
K_{\mathrm{AE}}(x,x')
=
\prod_{i=1}^{n}
\cos^2\left(\frac{x_i-x_i'}{2}\right),
\label{eq:angle_kernel}
\end{equation}

\noindent where $x$ and $x'$ represent two samples from the nominal training set and $n$ is the number of encoded features and corresponding qubits. The closed-form expression allows the angle-encoding kernel to be evaluated classically without quantum-circuit simulation. Angle encoding is therefore used primarily as an analytically tractable control to study global-measurement-induced concentration, rather than as a feature map intended to show quantum advantage.

\begin{figure*}
\centering \includegraphics[width=1.0\linewidth]{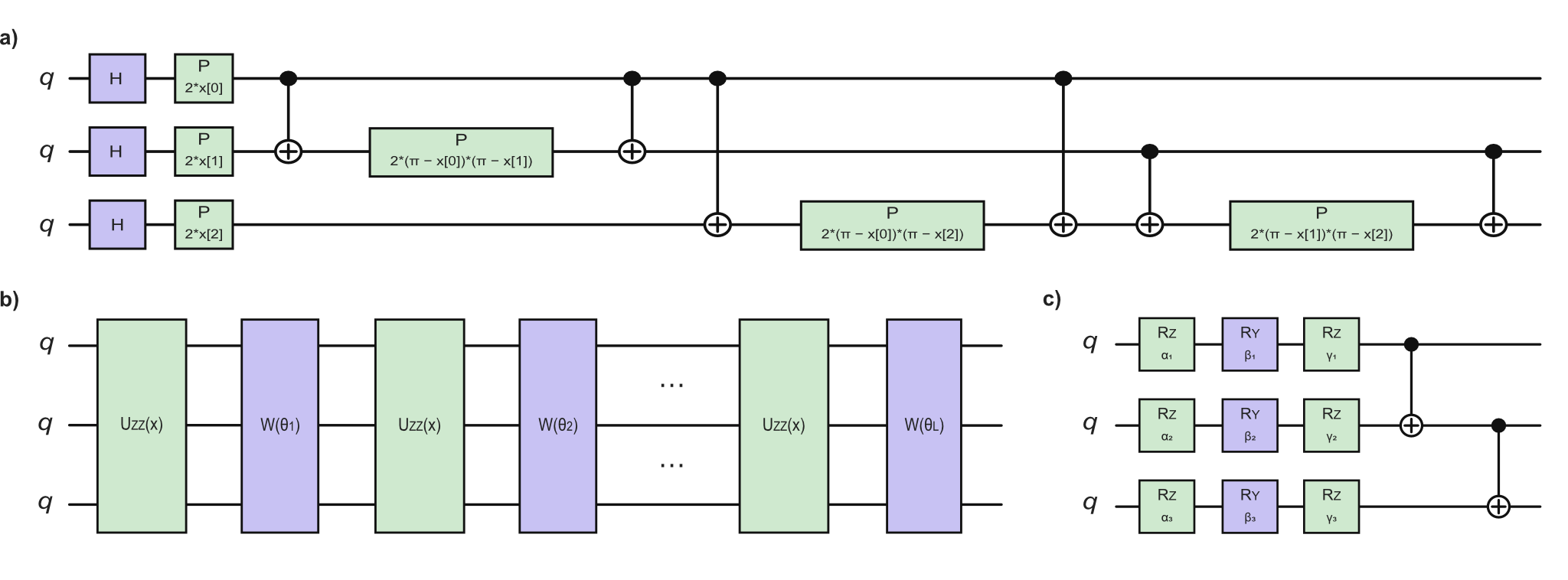} 
\caption{Quantum feature-map circuits used for kernel construction. 
(a) Example of the Full ZZ feature map for three qubits, consisting of an initial Hadamard layer followed by single-qubit data encoding and pairwise $ZZ$ interactions. 
(b) ZZ-R feature map, in which the Full ZZ encoding is repeated three times with fixed random unitary layers $\mathcal{R}_k$ inserted between successive data-encoding layers. 
(c) Structure of a fixed random layer $\mathcal{R}_k$, composed of single-qubit $R_Z$-$R_Y$-$R_Z$ rotations followed by nearest-neighbor CNOT entangling operations. The random rotation parameters are generated once and held fixed throughout kernel evaluation.}\label{fig:map} 
\end{figure*}

\subsubsection{Full $ZZ$ Feature Map}\label{sec:full-zz}
The Full $ZZ$ feature map provides the next level of expressivity considered in the concentration analysis. Unlike angle encoding, which encodes each feature independently, the $ZZ$ feature map additionally introduces data-dependent interactions between pairs of features. The Full $ZZ$ circuit is shown in Fig.~\ref{fig:map}(a) and consists of three components: an initial Hadamard layer, single-qubit phase encoding, and pairwise $ZZ$ interactions.

\paragraph{Hadamard layer.}
The feature map begins by applying a Hadamard gate to each of the $n$ qubits,

\begin{equation}
H^{\otimes n}|0\rangle^{\otimes n}
=
\frac{1}{\sqrt{2^n}}
\sum_{\mathbf{z}\in\{0,1\}^n}
|\mathbf{z}\rangle,
\label{eq:zz_hadamard}
\end{equation}

\noindent which prepares an equal superposition over all computational basis states. This allows subsequent data-dependent phase operations to assign different phases across the components of the superposition.

\paragraph{Single-feature phase encoding.}
Each classical feature $x_i$ is then encoded on its corresponding qubit using the phase gate

\begin{equation}
P(2x_i)
=
\begin{pmatrix}
1 & 0\\
0 & e^{i2x_i}
\end{pmatrix}.
\label{eq:zz_phase_gate}
\end{equation}

The phase gate is related to a rotation about the $z$-axis according to

\begin{equation}
P(2x_i)
=
e^{ix_i}R_Z(2x_i),
\qquad
R_Z(2x_i)
=
\exp(-ix_iZ).
\label{eq:phase_rz_relation}
\end{equation}

\noindent Because the factor $e^{ix_i}$ contributes only a global phase, the two operations produce equivalent encoded states for the purposes of the fidelity kernel. Thus, each feature controls the phase applied to its corresponding qubit.

\paragraph{Pairwise feature interactions.}
The distinguishing component of the $ZZ$ feature map is the introduction of interactions between pairs of features. In the canonical construction, the interaction between features $x_i$ and $x_j$ is determined by the data-mapping function

\begin{equation}
\phi_{ij}(\mathbf{x})
=
(\pi-x_i)(\pi-x_j)
\label{eq:zz_data_map}
\end{equation}

As shown in Fig.~\ref{fig:map}(a), this interaction can be implemented using a CNOT--phase--CNOT sequence,

\begin{equation}
\mathrm{CNOT}_{i\rightarrow j}\,
P_j\!\left(2\phi_{ij}(\mathbf{x})\right)\,
\mathrm{CNOT}_{i\rightarrow j}
\label{eq:zz_cnot_decomposition}
\end{equation}

\noindent which, up to a global phase, implements the two-qubit rotation

\begin{equation}
R_{ZZ}\!\left(2\phi_{ij}(\mathbf{x})\right)
=
\exp\left[
-i\phi_{ij}(\mathbf{x})Z_iZ_j
\right]
\label{eq:rzz_gate}
\end{equation}

The interaction phase therefore depends jointly on $x_i$ and $x_j$, introducing pairwise information that is absent when features are encoded using independent single-qubit rotations. In the full-connectivity construction, this operation is applied to every pair $i<j$.

Combining these three components, the canonical Full $ZZ$ feature-map transformation can be written, up to a global phase, as

\begin{equation}
\begin{split}
\mathcal{U}_{\mathrm{ZZ}}(\mathbf{x})
=
\left[
\prod_{i<j}
\exp\left(
-i(\pi-x_i)(\pi-x_j)Z_iZ_j
\right)
\right]\\
\left[
\prod_{i=1}^{n}
\exp\left(-ix_iZ_i\right)
\right]
H^{\otimes n}.
\end{split}
\label{eq:zz_encoding}
\end{equation}

This core $ZZ$ interaction structure is retained in this work, but the pairwise data-mapping function and the rotation angles are modified to reduce phase wrapping and preserve a larger dynamic range of the input features. Specifically, two changes are made to the canonical construction. First, the pairwise data-mapping function is defined as

\begin{equation}
\phi_{ij}(\mathbf{x}) = x_i x_j,
\label{eq:zz_data_map_modified}
\end{equation}

\noindent rather than $(\pi-x_i)(\pi-x_j)$. Second, single- and two-qubit rotations are applied at half the canonical angle, so that $R_Z(x_i)$ and $R_{ZZ}(x_i x_j)$ replace $R_Z(2x_i)$ and $R_{ZZ}(2\phi_{ij})$, respectively. Combining the Hadamard preparation, individual feature encoding, and pairwise interactions, the Full $ZZ$ transformation used throughout this work is therefore:

\begin{equation} 
\mathcal{U}_{\mathrm{ZZ}}(\mathbf{x})
=
\exp\!\left[
-\frac{i}{2}
\left(
\sum_i x_i Z_i
+
\sum_{i<j} x_i x_j Z_iZ_j
\right)
\right]
H^{\otimes n}.
\label{eq:zz_modified}
\end{equation}

Under the common scaling framework defined within section~\ref{sec:feature-map-selection} (Eq. \ref{eq:encoded-value}), these modifications reduce the sensitivity to phase wrapping and allow larger values of $\eta$ to be used for the $ZZ$ feature map. This limits the amount of input compression required while retaining more of the variation present within the nominal data.

\subsubsection{Full $ZZ$ Feature Map with Data Re-uploading}\label{sec:full-zz-reupload}
The final feature map, shown specifically in Fig. \ref{fig:map}(b), extends the Full $ZZ$ construction through data re-uploading~\cite{perezsalinas2020}, in which the input is encoded multiple times with intermediate quantum transformations applied between successive encoding layers.

The circuit consists of three Full $ZZ$ (Fig. \ref{fig:map}(a)) encoding layers separated by fixed random unitary transformations. Each intermediate transformation contains random single-qubit $R_Z$-$R_Y$-$R_Z$ rotations followed by a nearest-neighbour CNOT layer. The rotation parameters are randomly initialized once and held fixed throughout kernel evaluation, such that the increased circuit complexity does not introduce trainable parameters.

The complete three-layer transformation is

\begin{equation}
\mathcal{U}_{\mathrm{ZZ-DR}}(\mathbf{x}) =
\mathcal{U}_{\mathrm{ZZ}}(\mathbf{x})
\mathcal{R}_{2}
\mathcal{U}_{\mathrm{ZZ}}(\mathbf{x})
\mathcal{R}_{1}
\mathcal{U}_{\mathrm{ZZ}}(\mathbf{x}),
\label{eq:zz_reupload}
\end{equation}

\noindent where $\mathcal{U}_{\mathrm{ZZ}}(\mathbf{x})$ denotes the modified Full $ZZ$ encoding defined in Eq.~\ref{eq:zz_modified}, and $\mathcal{R}_{1}$ and $\mathcal{R}_{2}$ (shown in Fig. \ref{fig:map}(c)) denote the fixed random unitary transformations. The intermediate transformations alter the state produced by one encoding layer before the same input is re-encoded by the next, increasing the depth and expressivity of the resulting quantum embedding relative to the single-layer Full $ZZ$ feature map.

\section{Results}\label{results}
The results are presented in three phases. First, the downsampling experiment evaluates the sample efficiency of the proposed semi-supervised approach, demonstrating that high average precision can be achieved using relatively small nominal training sets. Then, we analyze the resulting kernel matrices to characterize kernel concentration. This analysis provides insight into the feasibility of each kernel on real quantum hardware. Finally, we implement the Nyström approximation (described in Sec. \ref{sec:nystrom-method}) as a means of further reducing the number of  quantum circuit evaluations required for train kernel construction. 

As described in the methodology  (Sec. \ref{sec:methodolgy}), a PCA sweep from 1 to 20 components was performed, corresponding to quantum kernels ranging from 1 to 20 qubits. Three quantum feature maps were evaluated: angle encoding (AE), the Full $ZZ$ feature map (ZZ), and the three-layer $ZZ$ data re-uploading feature map (ZZ-R). A classical radial basis function (RBF) kernel was additionally evaluated as a control for comparison with the quantum kernels. 

Throughout this section, feature-map and anomaly-detector abbreviations are combined when referring to a specific configuration. For example, AE-OC-SVM denotes an angle-encoding kernel optimized using OC-SVM, while ZZ-R-QSVR denotes a $ZZ$ re-uploading kernel optimized using QSVR. The same notation is used for the classical control, such that RBF-OC-SVM denotes the RBF kernel optimized using OC-SVM.

\begin{table*}[t]
\centering
\caption{Average precision (12 seeds, 20 PCA components) as training-set
size grows under cluster downsampling. Rank order stabilizes at 2{,}000 samples: 
RBF overtakes ZZ-R in OC-SVM between 1{,}000 and 2{,}000, after which the ordering is 
preserved through 4{,}000. $^\dagger$ denotes the classical RBF baseline.}
\label{tab:invariance}
\vspace{2mm}
\small
\renewcommand{\arraystretch}{1.4}
\setlength{\tabcolsep}{8pt}
\begin{tabular}{ll cccc c}
\toprule[0.9pt]
Kernel & Detector & AP@500 & AP@1000 & AP@2000 & AP@4000 & $\Delta$AP (4k$-$500)
\\
\midrule[0.4pt]
AE            & OC-SVM & 0.808 & 0.827 & 0.841 & 0.849 & 0.041 \\
RBF$^\dagger$ & OC-SVM & 0.789 & 0.815 & 0.830 & 0.842 & 0.053 \\
ZZ-R          & OC-SVM & 0.793 & 0.816 & 0.828 & 0.841 & 0.048 \\
ZZ-R          & QSVR  & 0.774 & 0.798 & 0.811 & 0.826 & 0.052 \\
AE            & QSVR  & 0.772 & 0.797 & 0.809 & 0.824 & 0.052 \\
ZZ            & OC-SVM & 0.763 & 0.795 & 0.805 & 0.820 & 0.057 \\
ZZ            & QSVR  & 0.758 & 0.784 & 0.797 & 0.812 & 0.054 \\
RBF$^\dagger$ & QSVR  & 0.757 & 0.782 & 0.792 & 0.806 & 0.049 \\
\bottomrule[0.9pt]
\end{tabular}
\end{table*}

\begin{table*}[t]
\centering
\caption{Power-law saturation analysis: AP$(N) = A - B/N^\alpha$. 
Two saturation thresholds are evaluated: $N^*(0.001)$ marks practical saturation 
(0.1\% performance gain per additional sample), while $N^*(0.0001)$ marks deep saturation 
(0.01\% gain/sample). All configurations exhibit sublinear convergence toward the fitted asymptote ($0<\alpha < 1$).}
\label{tab:powerlaw}
\vspace{2mm}
\small
\renewcommand{\arraystretch}{1.4}
\setlength{\tabcolsep}{8pt}
\begin{tabular}{ll cccc cc cc}
\toprule[0.9pt]
Kernel & Detector & $A$ & $B$ & $\alpha$ & $N^*(0.001)$ & $N^*(0.0001)$ & dAP/d$N$@4000 & $R^2$ \\
\midrule[0.4pt]
AE            & OC-SVM & 0.8670 & 2.14 & 0.5779 & 91   & 393  & 0.000003 & 0.9995 \\
RBF$^\dagger$ & OC-SVM & 0.8620 & 3.28 & 0.6124 & 112  & 466  & 0.000003 & 0.9992 \\
ZZ-R          & OC-SVM & 0.8657 & 1.65 & 0.5034 & 88   & 405  & 0.000003 & 0.9958 \\
ZZ-R          & QSVR  & 0.8620 & 1.17 & 0.4176 & 79   & 401  & 0.000004 & 0.9953 \\
AE            & QSVR  & 0.8528 & 1.56 & 0.4768 & 88   & 417  & 0.000004 & 0.9929 \\
ZZ            & OC-SVM & 0.8307 & 10.20 & 0.8083 & 146  & 523  & 0.000003 & 0.9857 \\
ZZ            & QSVR  & 0.8401 & 1.81 & 0.4984 & 94   & 436  & 0.000004 & 0.9945 \\
RBF$^\dagger$ & QSVR  & 0.8250 & 2.47 & 0.5792 & 100  & 428  & 0.000003 & 0.9890 \\
\bottomrule[0.9pt]
\end{tabular}
\end{table*}

\begin{table*}[t]
\centering
\caption{Performance at 500 nominal training samples relative to the fitted
asymptotic average precision $A$ obtained from the inverse power-law learning
curve. The gap from the fitted asymptote is calculated as
$(A-\mathrm{AP}_{500})/A \times 100\%$, while performance retained is
calculated as $\mathrm{AP}_{500}/A \times 100\%$. Across the evaluated
kernel-detector configurations, 500 training samples retain
89.79--93.19\% of the fitted asymptotic AP.}
\label{tab:asymptotic_gap}
\vspace{2mm}
\small
\renewcommand{\arraystretch}{1.4}
\setlength{\tabcolsep}{8pt}
\begin{tabular}{ll cccc}
\toprule[0.9pt]
\textbf{Kernel} & \textbf{Detector} & $\mathbf{AP@500}$ & $\mathbf{A}$ & \textbf{Gap from $A$ (\%)} & \textbf{Performance Retained (\%)} \\
\midrule[0.4pt]
AE            & OC-SVM & 0.808 & 0.8670 & 6.81  & 93.19 \\
RBF$^\dagger$ & OC-SVM & 0.789 & 0.8620 & 8.47  & 91.53 \\
ZZ-R          & OC-SVM & 0.793 & 0.8657 & 8.40  & 91.60 \\
ZZ-R          & QSVR  & 0.774 & 0.8620 & 10.21 & 89.79 \\
AE            & QSVR  & 0.772 & 0.8528 & 9.47  & 90.53 \\
ZZ            & OC-SVM & 0.763 & 0.8307 & 8.15  & 91.85 \\
ZZ            & QSVR  & 0.758 & 0.8401 & 9.77  & 90.23 \\
RBF$^\dagger$ & QSVR  & 0.757 & 0.8250 & 8.24  & 91.76 \\
\bottomrule[0.9pt]
\end{tabular}
\end{table*}

\subsection{Downsampling Sensitivity} \label{sec:downsample}
We first examine how anomaly-detection performance changes with the size of
the nominal training set. Across the 1--20 qubit PCA sweep, all evaluated
configurations achieve their highest mean average precision (AP) at 20
qubits. The downsampling analysis is therefore performed at this 20-qubit
operating point. As shown in Table~\ref{tab:invariance}, increasing the
training-set size improves mean AP for every evaluated kernel-detector
configuration; however, the magnitude of this improvement faces diminishing returns as additional nominal samples are introduced.

For example, AE-OC-SVM improves from $0.808$ AP at 500 samples to $0.827$ at
1{,}000 samples, corresponding to an increase of $0.019$. Increasing the
training set from 1{,}000 to 2{,}000 samples produces a smaller improvement
of $0.014$ AP, while doubling the training set again from 2{,}000 to
4{,}000 samples yields an improvement of only $0.008$ AP. Similar
diminishing returns are observed across the remaining configurations
(Table~\ref{tab:invariance}), indicating that larger nominal training sets
consistently improve performance within the experimentally evaluated range,
but provide progressively smaller incremental gains.

Not only does increasing the amount of nominal training data provide
diminishing returns in AP, but the relative performance of the evaluated
configurations also remains largely unchanged. The AP ranking changes only
once across the evaluated training-set sizes: between 1{,}000 and 2{,}000
samples, RBF-OC-SVM overtakes ZZ-R-OC-SVM
(Table~\ref{tab:invariance}). Beyond 2{,}000 samples, the ranking remains
unchanged through the largest training set considered. Thus, increasing the
nominal training-set size primarily improves absolute performance rather
than substantially altering the relative performance of the evaluated
kernel-detector configurations.

To quantify these diminishing returns and estimate the rate at which each
configuration approaches its limiting performance, the observed AP scaling
is subsequently modeled using a power-law saturation function \cite{figueroa2012predicting}.

\subsubsection{Power Law Saturation Model}\label{sec:downsampling-powerlaw}
To quantify the observed saturation behavior, the empirical mean AP measurements were fitted to a power-law saturation model,

\begin{equation}
\overline{\mathrm{AP}}(N) = A - \frac{B}{N^\alpha}
\end{equation}

\noindent where $N$ is the number of nominal training samples, \(\overline{\mathrm{AP}}(N)\) is the mean average precision across random seeds, $A$ is the asymptotic AP, and $B$ and $\alpha$ control the magnitude and rate of convergence toward the asymptote. The fitted parameters are summarized in Table~\ref{tab:powerlaw}. The model provides a close description of the observed scaling behavior, with $R^2$ values ranging from 0.986 to 0.9995 across the eight kernel-detector configurations. All fitted exponents satisfy $0 < \alpha < 1$, indicating sublinear convergence towards the fitted asymptotic performance. Because the marginal improvement

\begin{equation}
    \frac{B\alpha}{N^{\alpha+1}}
\end{equation}

\noindent decreases with N for $B>0$ and $\alpha>0$, the fitted models provide quantitative evidence of diminishing returns as the size of the nominal training set increases.

\subsubsection{Model Asymptotes and Convergence}\label{sec:power-law-saturation-results-analyze}

The fitted asymptotic AP values, $A$, range from 0.8250 to 0.8670
(Table~\ref{tab:powerlaw}), indicating meaningful differences in the
estimated limiting performance of the evaluated configurations. The fitted
convergence exponents, $\alpha$, vary from 0.4176 to 0.8083,
demonstrating that the rate at which performance approaches these asymptotes
also differs across configurations. Importantly,
the variation in $\alpha$ cannot be solely attributed to the kernel
representation or the anomaly-detection method, but instead depends on the
combination used in each configuration. For example, the ZZ feature map yields
$\alpha=0.4984$ when paired with QSVR, compared with $\alpha=0.8083$ when
paired with OC-SVM. Thus, the performance saturation with increasing
training-set size is a property of the complete kernel-method configuration
rather than each component independently.

The fitted power-law saturation models were also used to predict the number of training samples required for each kernel-detector combination to reach diminishing returns. We define $N(0.001)$ as the training-set size where the predicted improvement falls below 0.001 AP per additional sample, and $N(0.0001)$ as the training-set at where the predicted improvement falls below 0.0001 AP per additional sample. Across the evaluated configurations, $N(0.001)$ ranges from 79 to 146 samples, while the more stringent $N(0.0001)$ threshold ranges from 393 to 523 samples. Because the smallest experimentally evaluated training set contains 500 samples, these thresholds represent model-based extrapolations rather than directly observed saturation points. Nevertheless, the predicted diminishing returns are consistent with the experimentally observed scaling behavior, in which successive increases in the size of the training set produce progressively smaller improvements in AP.

The 500-sample results provide an experimentally evaluated reference point near this predicted diminishing-returns regime. At this training-set size, the evaluated configurations achieve AP values ranging from 0.757 to 0.808 and retain approximately 89.79--93.19\% of their fitted asymptotic performance. Thus, although additional nominal samples continue to improve performance, the majority of the predicted limiting performance is already achieved using only 500 training samples.

\subsection{Kernel Concentration Analysis}\label{sec:kernel-concentration-analysis}
The preceding analysis demonstrated that the nominal training set can be substantially downsampled while retaining most of the fitted asymptotic anomaly-detection performance, thus reducing the number of required kernel evaluations. However, hardware feasibility also depends on the number of measurement shots required to accurately estimate each kernel value. Therefore, this section examines kernel concentration as the number of qubits increases from 1 to 20. 

Since the power-law analysis in Sec. \ref{sec:power-law-saturation-results-analyze} indicates that 500 nominal samples already lie within or near the strong diminishing-return regime across the evaluated configurations, the concentration analysis is performed using the 500-sample training sets. This allows the shot requirements of the kernels to be evaluated within the reduced-sample regime targeted for hardware implementation. 

As established in Sec.~\ref{sec:concentration}, the experimental design controls the concentration induced by noise and entanglement, leaving global measurement and the expressivity of the feature map as the two concentration mechanisms considered in this analysis. First, we examine whether concentration behavior arises from the global fidelity measurement (Sec. \ref{sec:global-measurement-concentration}); then we characterize the relationship between feature-map expressivity and concentration (Sec. \ref{sec:feature-map-dependence}).
\begin{figure*}
\centering \includegraphics[width=1.0\linewidth]{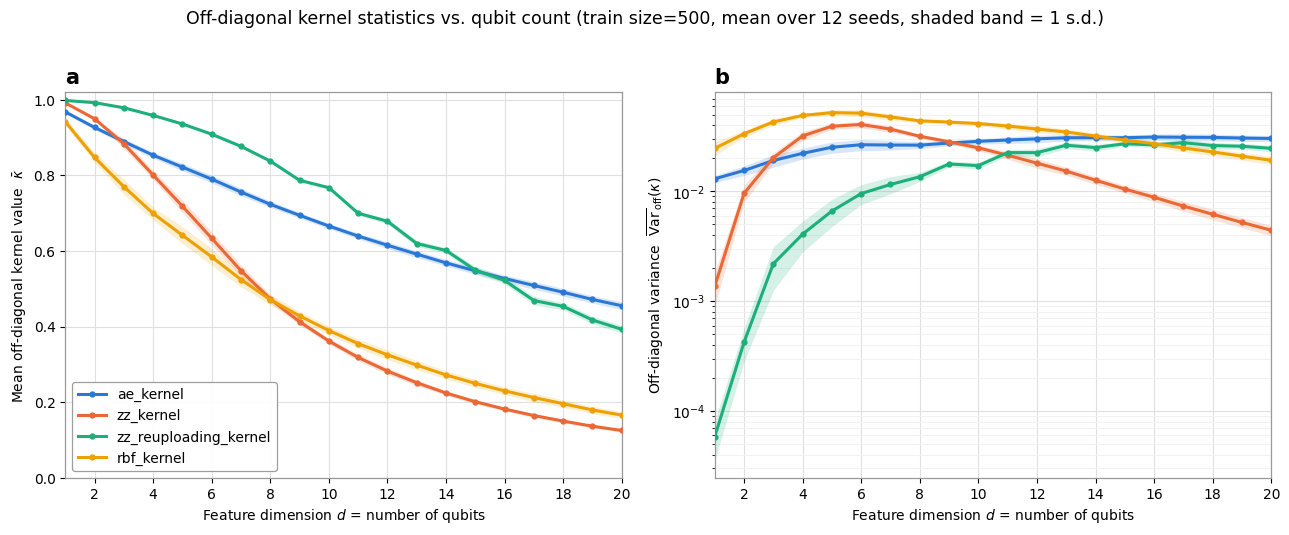} 
\caption{Kernel concentration statistics across the 1--20 qubit sweep for the evaluated quantum feature maps. (a) Median off-diagonal kernel value as a function of qubit count, characterizing the average pairwise fidelity between distinct samples. (b) Mean off-diagonal kernel variance as a function of qubit count, characterizing the spread of pairwise fidelities within each kernel matrix. Statistics are computed independently for each of the 12 sampled training sets at each qubit count and then averaged across seeds.}
\label{fig:kernel_concentration_sweep} 
\end{figure*}

\subsubsection{Global-Measurement Concentration}\label{sec:global-measurement-concentration}
To characterize global-measurement-induced concentration, we analyze the angle-encoding feature map as the number of qubits increases. Its shallow, non-entangling structure provides a low-expressivity setting, minimizing the contribution of expressivity-induced concentration and allowing the effects of global measurement to be more clearly examined. At each qubit count, we analyze the distribution of off-diagonal kernel matrix values using the median off-diagonal fidelity, $p_{50}$, and the off-diagonal variance. The median is used to determine whether typical off-diagonal fidelities become increasingly small with qubit count, while the variance measures whether their distribution becomes increasingly narrow. Together, these statistics characterize whether the kernel progresses toward the concentrated, identity-like regime described in Sec.~\ref{sec:concentration}.

Figure~\ref{fig:kernel_concentration_sweep} shows angle encoding statistics as the number of qubits increases from 1 to 20. Both quantities are independently evaluated for each of the 12 sampled training sets, and the reported values are averaged across seeds. Across this range, the median off-diagonal fidelity decreases from $0.998$ at one qubit to $0.458$ at 20 qubits, while the mean decreases from $0.968$ to $0.455$. As described in Sec.~\ref{sec:concentration}, product-state fidelity kernels can exhibit exponentially concentrated off-diagonal fidelities on the order of $10^{-6}$ at 20 qubits under the assumptions considered in Ref.~\cite{thanasilp2024}. 

Although the data distribution does not satisfy the same assumptions as this theoretical derivation, the $10^{-6}$ scale provides a useful reference to visualize the magnitude associated with severe exponential concentration. In comparison, the observed median fidelity of $0.458$ at 20 qubits remains more than five orders of magnitude above this reference scale. Thus, although pairwise fidelity decreases as the number of qubits increases, the angle-encoding kernel remains far from the severe concentration regime over the investigated range. The corresponding off-diagonal variance at 20 qubits remains $0.0303$, further indicating that the distribution has not collapsed tightly around an exponentially small value.

These results indicate that severe global-measurement-induced concentration does not emerge for the angle-encoding kernel within the investigated range of up to 20 qubits. Thanasilp \textit{et al.}~\cite{thanasilp2024} show that global-measurement-induced concentration can be mitigated when encoded data remain sufficiently localized in quantum Hilbert space. We hypothesize that the absence of severe concentration observed here may be partially attributed to the use of representative nominal training data whose proximity is preserved by the angle-encoding feature map. However, the present analysis does not directly establish this mechanism, and further investigation of the encoded-state geometry would be required to determine whether Hilbert-space localization is responsible for the observed behavior.

\begin{figure*}
    \centering
    \includegraphics[width=1.0\linewidth]{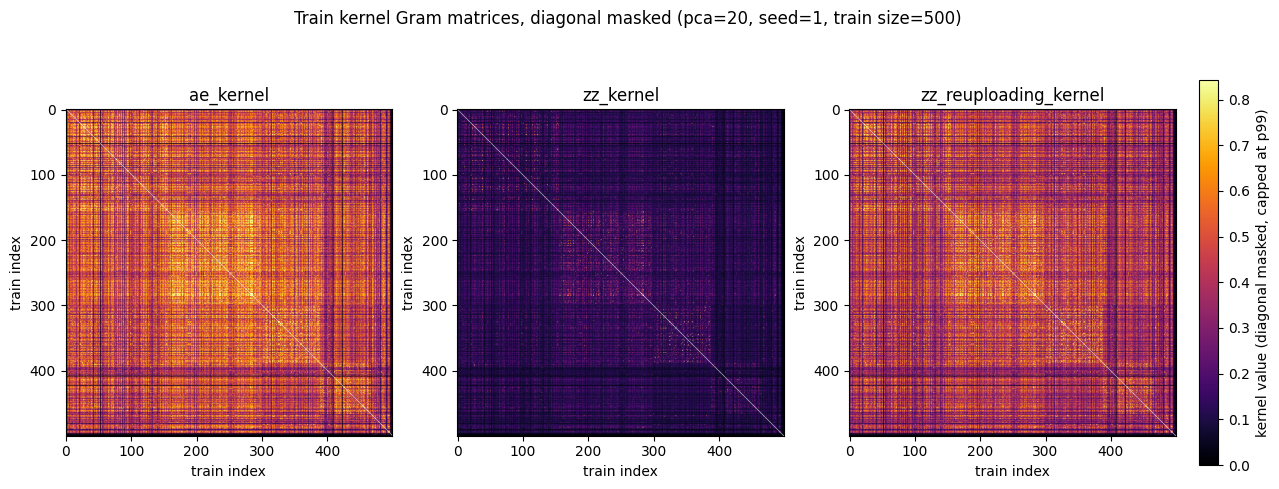}
    \caption{
    Fidelity Gram matrices for the 20-qubit Angle Encoding (AE), Full ZZ,
    and ZZ re-uploading (ZZ-R) quantum feature maps constructed from the
    same 500-sample nominal training set. Each pixel represents a pairwise
    kernel value $K(\mathbf{x}_i,\mathbf{x}_j)$ between two training samples,
    with the main diagonal equal to unity because each quantum state has
    unit fidelity with itself. The off-diagonal entries therefore represent
    the pairwise fidelities between distinct nominal samples and form the
    population used to calculate the kernel-concentration statistics reported
    in this work. Specifically, the off-diagonal median ($p_{50}$) is the
    median of all off-diagonal values visible in each heat map, while the
    off-diagonal variance measures the spread of these same values about
    their mean. Consequently, a fidelity kernel undergoing severe
    identity-like concentration would appear increasingly dominated by the
    unit diagonal as the off-diagonal entries become uniformly small and
    their variance decreases (this behavior is starting to be seen in the ZZ feature map (b)). The three matrices illustrate the distinct
    kernel geometries produced by the evaluated feature maps at the common
    20-qubit operating point.
    }
    \label{fig:gram_matrix}
\end{figure*}

\subsubsection{Feature-Map Dependent Concentration}\label{sec:feature-map-dependence}

The concentration behavior differs substantially across the three quantum feature maps: angle encoding, ZZ, and ZZ re-uploading. Figure~\ref{fig:gram_matrix} shows that the standard ZZ feature map (Sec. \ref{sec:full-zz}) undergoes a considerably stronger reduction in off-diagonal similarity than angle encoding. Its median off-diagonal fidelity decreases from $1.000$ at one qubit to $0.112$ at 20 qubits. The within-matrix off-diagonal variance at 20 qubits is $0.00442$, which is substantially smaller than the $0.0303$ observed for angle encoding. Together, the smaller off-diagonal values and reduced variance indicate stronger concentration of the ZZ kernel. 

\begin{table}[h]
\centering
\caption{Off-diagonal kernel statistics at the 20-qubit operating point. Values are aggregated across 12 random seeds.}
\label{tab:kernel_concentration}
\vspace{2mm}
\small
\renewcommand{\arraystretch}{1.4}
\setlength{\tabcolsep}{8pt}
\begin{tabular}{lccc}
\toprule[0.9pt]
\textbf{Feature Map} 
& $\bar{\kappa}$ 
& $p_{50}$ 
& $\operatorname{Var}_{\mathrm{off}}(\kappa)$ \\
\midrule[0.4pt]
Angle Encoding  & 0.455 & 0.458 & 0.0303  \\
ZZ              & 0.125 & 0.112 & 0.00442 \\
ZZ Re-uploading & 0.393 & 0.385 & 0.0246  \\
\bottomrule[0.9pt]
\end{tabular}
\end{table}

ZZ re-uploading (Fig. \ref{fig:gram_matrix}b) exhibits a markedly different concentration trend despite
using a deeper circuit with repeated data encoding and entangling operations.
At 20 qubits, its median off-diagonal fidelity remains $0.385$, compared with
only $0.112$ for the standard ZZ feature map (Fig. \ref{fig:gram_matrix}c). Its off-diagonal variance also
remains substantially larger than that of the standard ZZ feature map
(Table~\ref{tab:kernel_concentration}), indicating that the kernel values
retain greater spread. Together, these results show that ZZ re-uploading is
substantially less concentrated than the standard ZZ feature map at the
20-qubit operating point, despite its greater circuit depth and repeated
encoding structure.

The comparison between the ZZ feature map and ZZ Re-uploading indicates that increased feature-map expressivity does not necessarily correspond to stronger concentration for the data considered here. Although expressivity is a known source of severe exponential concentration, the deeper ZZ re-uploading architecture, which increases expressive capacity through repeated data encoding and additional circuit depth, produces substantially larger off-diagonal fidelities than the standard ZZ feature map.

The larger fidelities produced by angle encoding (Fig. \ref{fig:gram_matrix}a) and ZZ re-uploading indicate that their encoded nominal samples retain greater pairwise overlap than those produced by the standard ZZ feature map. Thus, other factors,
including the interaction between the feature map and the structure of the nominal data, appear to play an important role in determining the observed concentration behavior. While these results do not isolate the mechanism responsible for the difference between the two ZZ architectures, they
demonstrate that increasing the expressive capacity of the feature map does not necessarily produce stronger concentration.

Although the standard ZZ feature map exhibits substantially stronger concentration than angle encoding and ZZ re-uploading, its off-diagonal fidelities remain far from the exponentially small regime associated with severe fidelity-kernel concentration. At the 20-qubit operating point, the ZZ kernel retains a mean off-diagonal fidelity of $0.125$ and a median of $0.112$, both several orders of magnitude above the exponentially small reference scale discussed in Sec.~\ref{sec:concentration}. Angle encoding and ZZ re-uploading retain even larger off-diagonal fidelities. Thus, while the degree of concentration varies considerably across the evaluated feature maps, none enters a severe exponentially concentrated regime over the investigated range of up to 20 qubits.

Together, these results complement the preceding downsampling analysis.
Downsampling reduces the number of nominal samples required to maintain
competitive anomaly-detection performance, while the present analysis shows
that the resulting quantum kernels remain outside the severe concentration
regime over the evaluated dimensions, avoiding the exponentially increasing
shot requirements associated with resolving severely concentrated kernel
values. Having addressed both the number of training samples and the
measurement cost of individual kernel evaluations, we next investigate
whether the computational complexity of Gram-matrix construction can be
reduced from quadratic to linear in the number of training samples through
Nystr\"om approximation. 

\begin{table*}[t]
\centering
\caption{Nyström approximation performance: Average precision across landmark budgets. Columns represent the number of landmark points used for kernel matrix approximation, with the exact-kernel matrix computed using all 500 samples. Performance is generally stable across landmark selections, with most kernel-detector combinations achieving an average precision of at least $0.70$ using only 5 landmarks.}
\label{tab:nystrom_approximation}
\vspace{2mm}
\small
\renewcommand{\arraystretch}{1.4}
\setlength{\tabcolsep}{8pt}
\begin{tabular}{ll cccccccc}
\toprule[0.9pt]
Kernel & Method & 5 & 10 & 25 & 50 & 100 & 200 & 350 & Exact (500) \\
\midrule[0.4pt]
\multirow{2}{*}{AE} & OC-SVM & 0.694 & 0.546 & 0.605 & 0.663 & 0.668 & 0.775 & 0.799 & 0.808 \\
& QSVR & 0.727 & 0.715 & 0.731 & 0.745 & 0.745 & 0.761 & 0.768 & 0.772 \\
\midrule[0.4pt]
\multirow{2}{*}{ZZ-R} & OC-SVM & 0.724 & 0.731 & 0.730 & 0.753 & 0.760 & 0.765 & 0.779 & 0.793 \\
& QSVR & 0.744 & 0.740 & 0.754 & 0.772 & 0.771 & 0.774 & 0.777 & 0.774 \\
\midrule[0.4pt]
\multirow{2}{*}{RBF} & OC-SVM & 0.696 & 0.659 & 0.524 & 0.443 & 0.509 & 0.745 & 0.782 & 0.789 \\
& QSVR & 0.736 & 0.730 & 0.737 & 0.739 & 0.740 & 0.747 & 0.740 & 0.757 \\
\midrule[0.4pt]
\multirow{2}{*}{ZZ} & OC-SVM & 0.724 & 0.734 & 0.737 & 0.742 & 0.743 & 0.754 & 0.747 & 0.763 \\
& QSVR & 0.737 & 0.728 & 0.733 & 0.738 & 0.740 & 0.748 & 0.698 & 0.758 \\
\bottomrule[0.9pt]
\end{tabular}
\end{table*}

\subsection{Nyström Approximation}\label{sec:nystrom-approx-results}
The Nystr\"om approximation provides an additional mechanism for reducing
the number of quantum kernel evaluations required for Gram-matrix
construction (Table~\ref{tab:nystrom_approximation}). Using $m$ landmark
points reduces the kernel-evaluation complexity from $\mathcal{O}(N^2)$ to
$\mathcal{O}(Nm)$, where $m \ll N$. The analysis is performed using the
500-sample Gram matrices considered in the preceding concentration analysis.
Under the most aggressive approximation, only five landmarks are used,
corresponding to $1\%$ of the training set.

The evaluated configurations retain substantial anomaly-detection
performance even at small landmark budgets, although their robustness to
approximation differs. QSVR is particularly robust: with only five
landmarks, AP ranges from $0.727$ to $0.744$, compared with
$0.757$--$0.774$ for the corresponding exact kernels, retaining
approximately $94$--$97\%$ of exact-kernel performance. OC-SVM is more
sensitive to aggressive approximation, and the magnitude of performance reduction varies across feature maps. Increasing the landmark budget generally recovers performance toward the exact-kernel result, although this
convergence is not strictly monotonic for every configuration.

These results show that robustness to Nystr\"om approximation is determined
by the combination of feature map and anomaly-detection method rather than
either component independently. Even for the same feature map, QSVR and
OC-SVM exhibit different responses to reductions in the landmark budget,
with QSVR generally retaining a larger fraction of its exact-kernel AP.

These results demonstrate that the Nyström approximation provides a complementary mechanism to training-set downsampling  for reducing the cost of Gram matrix construction. Downsampling reduces the number of training samples used to construct the Gram matrix, while the Nyström approximation reduces the computational complexity. Together, the results indicate that substantial reductions in quantum kernel training requirements are possible while maintaining competitive anomaly-detection performance.

\section{Conclusion}
This work demonstrates that non-variational quantum kernel anomaly detectors can be trained with substantially reduced resource requirements by addressing three major sources of cost: the number of samples used to construct the quantum Gram matrix, the measurement cost of estimating individual kernel values, and the number of kernel evaluations required for Gram-matrix construction. Across the evaluated feature maps and anomaly-detection methods, high average precision is retained using only a small fraction of the available nominal data.

First, the training-set analysis revealed strong diminishing returns as the number of nominal samples increased. The fitted power-law models predict that the marginal improvement falls below $0.001$ AP per additional sample at training-set sizes of 79--146 samples and below $0.0001$ AP per additional sample at 393--523 samples. At the experimentally evaluated 500-sample operating point, the configurations retain approximately 89.79--93.19\% of their fitted asymptotic AP.

Second, the kernel-concentration analysis demonstrates that this reduced-sample operating regime does not encounter severe exponential concentration over the investigated range of 1--20 qubits. Although all three quantum feature maps exhibit decreasing off-diagonal fidelities as the number of qubits increases (Fig. \ref{fig:kernel_concentration_sweep}), no feature map enters the exponentially small regime associated with severely concentrated kernels. The standard ZZ feature map exhibits the strongest concentration, whereas angle encoding and ZZ re-uploading retain substantially larger off-diagonal fidelities. Notably, ZZ re-uploading remains less concentrated than standard ZZ despite its greater circuit depth and repeated data encoding, indicating that stronger concentration does not simply follow from increasing feature-map complexity. Consequently, none of the evaluated kernels enters the regime in which exponentially small fidelities would require prohibitively large measurement budgets to resolve on quantum hardware.

Finally, Nystr\"om approximation further reduces the number of quantum kernel evaluations required during training. For a fixed landmark budget $m \ll N$, Nystr\"om reduces the kernel-evaluation complexity from $\mathcal{O}(N^2)$ to $\mathcal{O}(Nm)$, allowing the cost of an already downsampled training set to be reduced further. With only five landmarks and 500 nominal samples, QSVR retains approximately 94--97\% of its exact-kernel AP across the evaluated kernels. However, the response to these resource reductions is configuration-dependent. OC-SVM generally achieves stronger performance under direct training-set downsampling, whereas QSVR is particularly robust to aggressive Nystr\"om approximation. Similarly, the fitted convergence exponent $\alpha$ varies with the combination of feature map and anomaly-detection method. Resource-efficient quantum kernel design therefore depends on the complete kernel-detector configuration rather than on either component in isolation.

The practical magnitude of these resource reductions can be illustrated
using the ZZ-QSVR configuration and the largest training set evaluated in this work. For $N=4{,}000$, the observed AP is $0.812$. Since fidelity kernels are symmetric and their diagonal entries are known to equal one, constructing the exact kernel requires only the unique off-diagonal entries, corresponding to 7{,}998{,}000 quantum kernel evaluations. At IonQ's minimum estimated cost of \$168 per kernel evaluation~\cite{ionq2025}, this corresponds to an illustrative training cost of approximately \$1.34 billion. Downsampling to the experimentally evaluated 500-sample operating point reduces the exact kernel to 124{,}750 evaluations, corresponding to approximately \$21.0 million under the same pricing assumption. Applying a five-landmark Nystr\"om approximation further reduces this requirement to approximately 2{,}485 evaluations, corresponding to an illustrative cost of \$417{,}480. Overall, reducing the experimentally evaluated 4{,}000-sample kernel to the 500-sample, five-landmark Nystr\"om approximation decreases the number of required quantum kernel evaluations by approximately $99.97\%$, or more than 3{,}200-fold. For ZZ-QSVR, this reduction decreases the observed AP from $0.812$ to $0.737$, corresponding to a relative reduction of approximately $9.2\%$. Moreover, the fitted power-law model predicts an asymptotic AP of only $A=0.8401$ for ZZ-QSVR. Thus, even relative to the predicted infinite-sample limit, the aggressively reduced model remains within approximately $12.3\%$ of the fitted maximum AP. The resulting AP of $0.737$ also remains substantially above the $0.05$ random-ranking baseline associated with the $5\%$ anomaly prevalence of the test set. Taken together, these results demonstrate that nominal-data downsampling, favorable kernel-concentration behavior, and low-rank approximation can reduce the resource requirements associated with training non-variational
quantum kernel anomaly-detection models by several orders of magnitude while
retaining meaningful anomaly-discrimination performance.

\section*{Acknowledgments}
Kayla J. Rodriguez acknowledges salary support from the IBM-HBCU Quantum Center and fellowship support from the National Science Foundation Research Traineeship (NSF NRT) program. We would like to acknowledge Christopher Sherald for assisting with the literature review on the Nystr\"om approximation.

\section*{AI Acknowledgement}
The authors used Claude (Anthropic) to assist with editing and formatting tables and with identifying relevant references. All AI-assisted content was reviewed and verified by the authors, who take full responsibility for the content of this work.

\clearpage
\bibliography{references}
\bibliographystyle{IEEEtran}

\clearpage
\appendix

\section{One-Class Kernel Anomaly Detection}
\label{sec:one_class_anomalies}

This appendix provides an overview of the One-Class Support Vector Machines (OC-SVMs) used in this work. The first part, Sec.~\ref{sec:OC-SVM}, provides a general overview of OC-SVMs. The following two parts, Sec.~\ref{sec:support-vector-data} and Sec.~\ref{sec:SVR}, give an overview of the specific OC-SVM models used in this work.

\subsection{One-Class Support Vector Machines}
\label{sec:OC-SVM}

\begin{figure}
\centering
\includegraphics[width=0.95\linewidth]{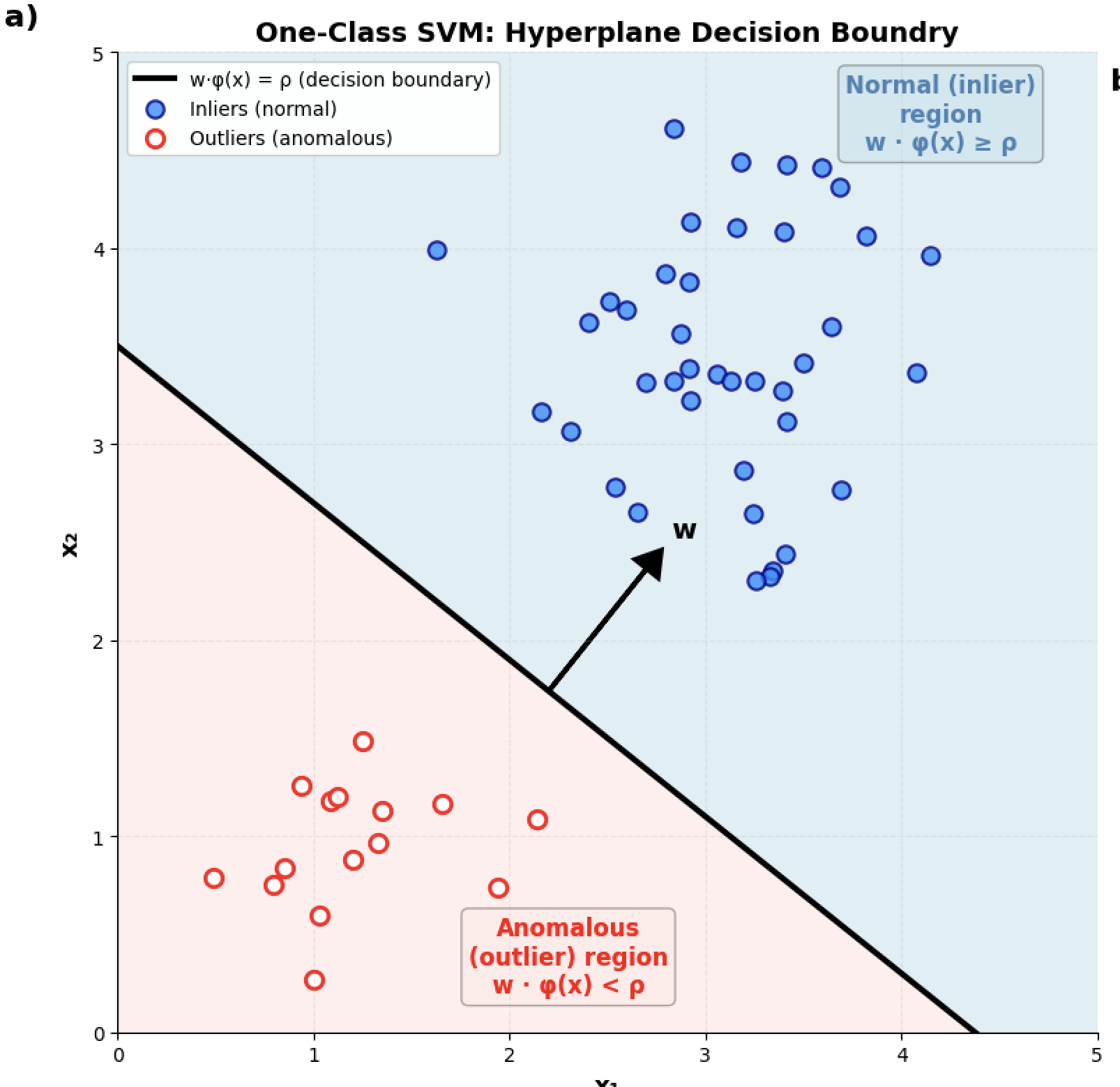}
\caption{Support Vector Methods for anomaly detection and regression. 
\textbf{(a)} One-Class SVM: A hyperplane decision boundary $\mathbf{w} \cdot \boldsymbol{\phi}(\mathbf{x}) = \rho$ separates the normal (inlier) region from anomalous (outlier) regions in the feature space. Support vectors (circled points) lie close to the decision boundary and determine the margin. The normal region is defined by $\mathbf{w} \cdot \boldsymbol{\phi}(\mathbf{x}) \geq \rho$, with anomalies falling in the complementary region $\mathbf{w} \cdot \boldsymbol{\phi}(\mathbf{x}) < \rho$.
}
\label{fig:SVM_Methods}
\end{figure}

The One-Class Support Vector Machine (OC-SVM)~\cite{scholkopf1999} is a kernel-based anomaly detection method that maps nominal data into a feature space induced by a kernel and learns a maximum-margin hyperplane that separates the mapped data from the origin, as illustrated in Fig.~\ref{fig:SVM_Methods}. Points on the origin side of the learned decision boundary are classified as anomalies. The dual optimization problem is

\begin{equation}
\begin{aligned}
\max_{\alpha}\quad
&-\frac{1}{2}\sum_{i,j=1}^{m}
\alpha_i\alpha_jK(x_i,x_j)\\
\text{s.t.}\quad
&0\leq\alpha_i\leq\frac{1}{\nu m},\\
&\sum_{i=1}^{m}\alpha_i=1.
\end{aligned}
\label{eq:ocsvm_dual}
\end{equation}

Here, $m$ denotes the number of training samples, $\alpha_i$ are the Lagrange multipliers associated with each training sample, and $K(x_i,x_j)$ is the kernel function evaluated between training samples $x_i$ and $x_j$. Training samples with nonzero $\alpha_i$ are called \textit{support vectors}. The parameter $\nu\in(0,1]$ controls the trade-off
between the fraction of training samples allowed to lie outside the
learned decision boundary and the number of support vectors~\cite{scholkopf1999}. 

\subsection{Support Vector Data Description}\label{sec:support-vector-data}
Support Vector Data Description (SVDD)~\cite{svdd04} formulates one-class anomaly detection as the fitting of a hypersphere in the kernel-induced feature space. The hypersphere  encloses the normal training data as illustrated in Fig.~\ref{fig:svdd}, and points outside the learned hypersphere are classified as anomalies. The center of the learned hypersphere is given by a linear combination
of the mapped training samples,
\begin{equation}
c = \sum_{i=1}^{m}\alpha_i\Phi(x_i),
\end{equation}
where again, $\alpha_i$ are the Lagrange multipliers associated with the
training samples, and samples with nonzero $\alpha_i$ are called support
vectors. The squared distance of a sample $x$ from the hypersphere center can again be expressed entirely in terms of kernel evaluations as
\begin{equation}
\begin{aligned}
\|\Phi(x)-c\|^2
=
&\,K(x,x)
-2\sum_{i=1}^{m}\alpha_i K(x,x_i)\\
&+\sum_{i,j=1}^{m}\alpha_i\alpha_j K(x_i,x_j).
\end{aligned}
\label{eq:svdd_score}
\end{equation}
A sample is classified as anomalous when its squared distance from the hypersphere center exceeds the learned radius,
\begin{equation}
\|\Phi(x)-c\|^2 > R^2,
\label{eq:svdd_decision}
\end{equation}
where the radius $R$ is determined during training from the support vectors. As previously stated, by using a quantum kernel as defined in Eq.~\ref{eq:quantum_kernel} the SVDD fits a hyperplane in quantum Hilbert space to produce a quantum SVDD (QSVDD).

Although OC-SVM and SVDD use different geometric formulations, their decision boundaries are equivalent for kernels with a constant diagonal, i.e., $K(x,x)=\mathrm{const}$~\cite{svdd04}. The fidelity-based quantum kernels considered in this work satisfy this condition because they are constructed from normalized quantum states, giving $K(x,x)=1$. Consequently, the QOC-SVM and quantum SVDD formulations produce equivalent decision boundaries for all feature maps considered in this study. We therefore use the OC-SVM formulation throughout the experiments and do not treat quantum SVDD as a separate experimental method, but it is included because its visualization is more intuitive, and provides a good mental model when considering the results.

\subsection{Support Vector Regression}\label{sec:SVR}
\begin{figure}
\centering
\includegraphics[width=1.05\linewidth]{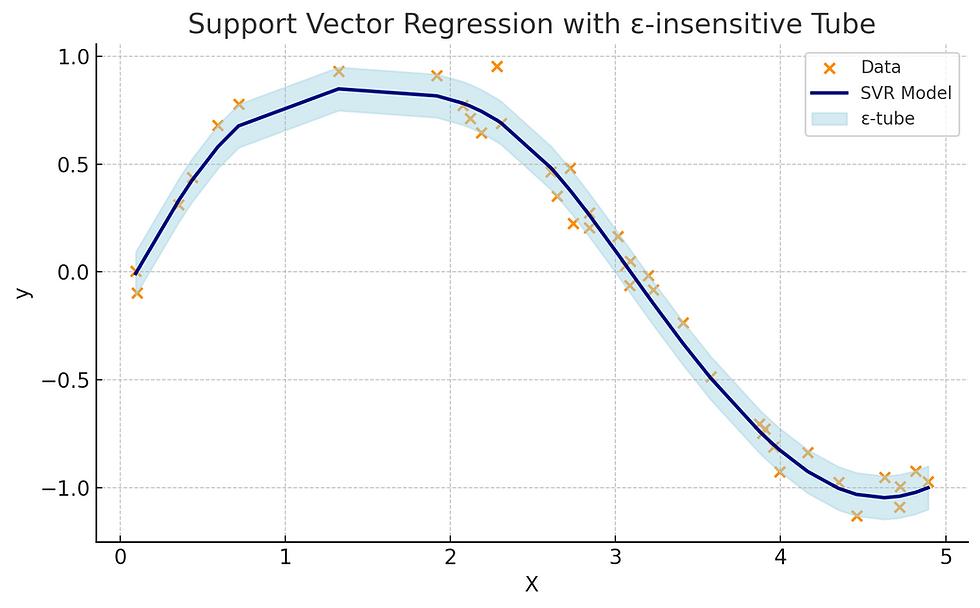}
\caption{Support Vector Regression (SVR): Predictions within an $\varepsilon$-insensitive tube of the regression function incur no loss, while deviations outside the tube incur an $\varepsilon$-insensitive loss. The parameter $C$ controls the penalty assigned to these deviations, with larger values placing greater emphasis on fitting the training data.
}
\label{fig:SVM_Methods-tube-insensitivity}
\end{figure}

Support Vector Regression (SVR)~\cite{drucker1996} is not an explicit anomaly detection method, but rather a regression method that learns a function by allowing predictions to deviate from training targets by a specified tolerance $\varepsilon$ (Figure \ref{fig:SVM_Methods-tube-insensitivity}). A parameter $C$ controls the trade-off between model complexity and the penalty assigned to deviations outside the $\varepsilon$-insensitive tube. The dual formulation of the SVR optimization problem is

\begin{equation}
\begin{aligned}
\max_{\alpha, \alpha^*} \quad
& -\frac{1}{2} \sum_{i,j=1}^{m}
(\alpha_i - \alpha_i^*)
(\alpha_j - \alpha_j^*)
K(x_i, x_j) \\
& - \varepsilon \sum_{i=1}^{m}
(\alpha_i + \alpha_i^*)
+ \sum_{i=1}^{m}
y_i(\alpha_i - \alpha_i^*) \\
\text{s.t.} \quad
& \sum_{i=1}^{m}
(\alpha_i - \alpha_i^*) = 0, \\
& 0 \leq \alpha_i, \alpha_i^* \leq C.
\end{aligned}
\label{eq:svr_dual}
\end{equation}

\noindent where $K(x_i,x_j)$ is the kernel function and $\alpha_i$ and $\alpha_i^*$ are the Lagrange multipliers associated with the lower and upper sides of the $\varepsilon$-insensitive tube respectively. The resulting regression function is

\begin{equation}
f(x) =
\sum_{i=1}^{m}(\alpha_i-\alpha_i^*)K(x_i,x)+b,
\label{eq:svr-decision-function}
\end{equation}

\noindent where $b$ is the learned bias term.

Even though SVR was not originally formulated for anomaly detection, it can be adapted to the task by training the regressor exclusively on samples from the nominal class and using its prediction error as an anomaly score. Because a standard SVR produces a single scalar output, a separate regressor is trained for each feature of the input. The $d$-th regressor $f_d$ is trained to reproduce the $d$-th feature, with target $y_i^{(d)}=(x_i)_d$. For a sample $x$, the outputs of the $D$ regressors are concatenated to form the reconstructed vector

\begin{equation}
\hat{x} =
\left[
f_1(x), f_2(x), \ldots, f_D(x)
\right].
\end{equation}

The reconstruction error $s(x)$ between the actual feature values and the predicted values and serves as the anomaly score for a sample $x$,

\begin{equation}
s(x)=\sum_{d=1}^{D}\big(x_d-\hat{x}_d\big)^2,
\label{eq:svr_score}
\end{equation}

where $D$ is the total number of features describing a sample. Nominal points, (whose structure is learned during training), are expected to have a lower reconstruction error than anomalous points, and consequentially a lower anomaly score. When a samples anomaly score $s(x)$ exceeds a certain threshold $\tau$, it is classified as anomalous. A sample is classified as anomalous when its anomaly score $s(x)$ exceeds a predefined threshold $\tau$. In this work, the threshold is set to three times the mean reconstruction error over the nominal training set:

\begin{equation}
\tau =
3\left(\frac{1}{m}\sum_{i=1}^{m}s(x_i)\right).
\end{equation}

Rather than simply assigning a binary class label, the reconstruction error provides a continuous measure of a sample's deviation from the learned nominal structure, which gives us a measure of \textit{how} anomalous a given sample is. Importantly, although $D$ independent regressors are required to reconstruct a $D$-dimensional input, they can all use the same kernel matrix constructed from the same nominal training samples. Each regressor uses this shared kernel matrix while learning to predict a different feature. Therefore, increasing the number of reconstructed features does not require additional quantum kernel evaluations or the construction of additional quantum kernel matrices. All additional cost is classical, and for our intents and purposes, free. 

Reconstruction-based anomaly detection has already been investigated in the quantum setting. Tscharke et al.~\cite{tscharke2023semisupervised} introduced a quantum-kernel version of this approach by replacing the classical kernel with a quantum kernel. They subsequently investigated the same QSVR-based anomaly detection framework on superconducting quantum hardware~\cite{tscharke2025nisq}.

\end{document}